\documentclass[preprint,prd,aps,showpacs,showkeys,nofootinbib]{revtex4}
\usepackage{graphicx}
\usepackage{dcolumn}
\usepackage{bm}
\usepackage{diagbox}
\usepackage{color}
\usepackage[dvipsnames]{xcolor}
\usepackage{amssymb}
\usepackage{makecell}
\usepackage{subfigure}
\usepackage{amsmath}

\definecolor{bright-blue}{RGB}{0,255,255}
\definecolor{light-gray}{gray}{0.88}
\definecolor{dark-gray}{gray}{0.40}
\begin{document}

\title{The flavor transition process $b\rightarrow s\gamma$ in the $U(1)_X$SSM with the mass insertion approximation}
\author{Xin-Xin Long$^{1,2,3}$, Shu-Min Zhao$^{1,2,3}$\footnote{zhaosm@hbu.edu.cn}, Xi Wang$^{1,2,3}$, Yi-Tong Wang$^{1,2,3}$, Tong-Tong
Wang$^{1,2,3}$, Jian-Bin Chen$^{4}$, Hai-Bin Zhang$^{1,2,3}$, Tai-Fu Feng$^{1,2,3,5}$}

\affiliation{$^1$ Department of Physics, Hebei University, Baoding 071002, China}
\affiliation{$^2$ Key Laboratory of High-precision Computation and Application of Quantum Field Theory of Hebei Province, Baoding 071002,
China}
\affiliation{$^3$ Research Center for Computational Physics of Hebei Province, Baoding 071002, China}
\affiliation{$^4$ College of Physics and Optoelectronic Engineering, Taiyuan University of Technology, Taiyuan 030024, China}
\affiliation{$^5$ Department of Physics, Chongqing University, Chongqing 401331, China}
\date{\today}

\begin{abstract}
People extend the MSSM with the local gauge group $U(1)_X$ to obtain the $U(1)_X$SSM. In the framework of the $U(1)_X$SSM, we study the
flavor transition process $b\rightarrow{s\gamma}$ with the mass insertion approximation (MIA). By the MIA method and some
reasonable parameter assumptions, we can intuitively find the parameters that have obvious effect on the analytic results of the flavor
transition process $b\rightarrow{s\gamma}$. By means of the influences of different sensitive parameters, we can obtain reasonable results
to better fit the experimental data.

\end{abstract}


\maketitle

\section{Introduction}

The rare decay $\bar B\to X_s\gamma$ is the Flavor Changing Neutral Current (FCNC) process with great research value, whose CP averaged
branching ratio is used to constrain many models of new physics. People have conducted multi-angle research on the  process. The authors
of Refs. \cite{bsy1,bsy2,bsy3} present the calculation of the rate of the inclusive decay $\bar B\to X_s\gamma$ in the two-Higgs doublet model (THDM).
The supersymmetric effect on $\bar B\to X_s\gamma$ is discussed in Refs. \cite{bsy4,bsy5,bsy6,bsy7,bsy8,bsy9}.

Since the influence of new physics on FCNC process is mainly derived from the loop diagrams process, the decay process $b\to
s\gamma$ is very sensitive to new physics beyond the Standard Model (SM). The average experimental data on the branching ratio of the
inclusive $\bar{B}\rightarrow X_s\gamma$ is \cite{Average}
\begin{eqnarray}
BR(\bar{B}\rightarrow X_s\gamma)_{exp}=(3.40\pm0.21)\times10^{-4},
\label{exp-bsg}
\end{eqnarray}
and the prediction of SM at next-next-to-leading order (NNLO) is \cite{SM1,SM2,SM3,SM4,SM5,SM6,SM7,SM8}
\begin{eqnarray}
BR(\bar{B}\rightarrow X_s\gamma)_{SM}=(3.36\pm0.23)\times10^{-4}.
\label{SM-bsg}
\end{eqnarray}

A famous extension of the SM is the minimal supersymmetric extension of the standard model (MSSM) \cite{MSSM}. As a popular extension, the MSSM
solves many problems. However, the explanation of neutrino oscillation requires tiny neutrino mass, which can not be naturally generated by the
MSSM. So people extend the MSSM into multiple models, among which, the $U(1)_X$ extension of the MSSM is called as the $U(1)_X$SSM
\cite{UX2,UX3} with the local gauge group $SU(3)_C\otimes SU(2)_L \otimes U(1)_Y \otimes U(1)_X$. Compared to the MSSM, the $U(1)_X$SSM has
more superfields such as right-handed neutrinos. The right-handed neutrinos and the added Higgs singlets can explain the tiny mass of neutrino. It can
alleviate the $\mu$ problem of the MSSM, because $\mu\hat{H}_u\hat{H}_d$ and $\lambda_H\hat{S}\hat{H}_u\hat{H}_d$ in the $U(1)_X$SSM produce the
effective $\mu_{H}'=\mu+\lambda_Hv_S/\sqrt{2}$. Mixing the CP-even parts of $H_d,~H_u,~\eta,~\bar{\eta},~S$ can improve the masses of the
tree-level lightest and sub-lightest CP-even Higgs particles.

The mass eigenstate basis is the most commonly used method in FCNC research. However, the mass eigenstate method depends on
the mass eigenvalues of particles and rotation matrices, which makes it difficult to find the sensitive parameters clearly and intuitively. To
solve this problem, we use another method - Mass Insertion Approximation (MIA) \cite{UX3,MIA4,MIA1,MIA2,MIA3}. The MIA
results can be alternatively obtained from expanding properly the expressions in the mass eigensate basis \cite{MEN}. At the analytical level,
it avoids tedious calculations and reduces the chances of errors. In addition, its relatively concise and clear analytical results are helpful
for the analysis of sensitive parameters.

To sum up, we investigate the FCNC process $b\rightarrow s\gamma$ under the $U(1)_X$SSM by MIA. Our paper is organized as
follows. In Sec.II, we mainly introduce the $U(1)_X$SSM including its superpotential and the general soft breaking terms. The analytic
expressions for $b\rightarrow s\gamma$ decay in the $U(1)_X$SSM are given in Sec.III. In Sec.IV, we give the numerical analysis. And our
conclusions are summarized in Sec.V.

\section{The $U(1)_X$SSM}

We extend the MSSM with the local gauge group $U(1)_X$ to obtain the $U(1)_X$SSM, whose local gauge group is $SU(3)_C\otimes SU(2)_L \otimes
U(1)_Y\otimes U(1)_X$. There are new superfields in it compared to the MSSM, including three Higgs singlets
$\hat{\eta},~\hat{\bar{\eta}},~\hat{S}$ and right-handed neutrinos $\hat{\nu}_i$. The $U(1)_X$SSM has several advantages than the MSSM. For
example, through the seesaw mechanism, light neutrinos obtain tiny masses at the tree level. The superpotential of this model is:
\begin{eqnarray}
&&W=l_W\hat{S}+\mu\hat{H}_u\hat{H}_d+M_S\hat{S}\hat{S}-Y_d\hat{d}\hat{q}\hat{H}_d-Y_e\hat{e}\hat{l}\hat{H}_d+\lambda_H\hat{S}\hat{H}_u\hat{H}_d
\nonumber\\&&~~~~~~+\lambda_C\hat{S}\hat{\eta}\hat{\bar{\eta}}+\frac{\kappa}{3}\hat{S}\hat{S}\hat{S}+Y_u\hat{u}\hat{q}\hat{H}_u+Y_X\hat{\nu}\hat{\bar{\eta}}\hat{\nu}
+Y_\nu\hat{\nu}\hat{l}\hat{H}_u.
\end{eqnarray}

We show the explicit forms of two Higgs doublets and three Higgs singlets here
\begin{eqnarray}
%%%%%%%%%%%%%%%%%%%%%%%%%%%%%%%%%%%%%%%%%%%%%%%%%%%%%%%%%%%%%%%%%%%%
&&H_{u}=\left(\begin{array}{c}H_{u}^+\\{1\over\sqrt{2}}\Big(v_{u}+H_{u}^0+iP_{u}^0\Big)\end{array}\right),
~~~~~~
H_{d}=\left(\begin{array}{c}{1\over\sqrt{2}}\Big(v_{d}+H_{d}^0+iP_{d}^0\Big)\\H_{d}^-\end{array}\right),
\nonumber\\
&&\eta={1\over\sqrt{2}}\Big(v_{\eta}+\phi_{\eta}^0+iP_{\eta}^0\Big),~~~~~~~~~~~~~~~
\bar{\eta}={1\over\sqrt{2}}\Big(v_{\bar{\eta}}+\phi_{\bar{\eta}}^0+iP_{\bar{\eta}}^0\Big),\nonumber\\&&
\hspace{4.0cm}S={1\over\sqrt{2}}\Big(v_{S}+\phi_{S}^0+iP_{S}^0\Big).
%%%%%%%%%%%%%%%%%%%%%%%%%%%%%%%%%%%%%%%%%%%%%%%%%%%%%%%%%%%%%%%%%%%%
\label{gauge2}
\end{eqnarray}
In Eq.(\ref{gauge2}), $v_{u},~v_{d},~v_\eta$,~ $v_{\bar\eta}$ and $v_S$  respectively represent the vacuum expectation values(VEVs) of the
Higgs superfields $H_u$, $H_d$, $\eta$, $\bar{\eta}$ and $S$. And two angles are defined as $\tan\beta=v_{u}/v_{d}$ and
$\tan\beta_\eta=v_{\bar{\eta}}/v_{\eta}$.

The soft SUSY breaking terms of this model are shown as
\begin{eqnarray}
&&\mathcal{L}_{soft}=\mathcal{L}_{soft}^{MSSM}-B_SS^2-L_SS-\frac{T_\kappa}{3}S^3-T_{\lambda_C}S\eta\bar{\eta}
+\epsilon_{ij}T_{\lambda_H}SH_d^iH_u^j\nonumber\\&&\hspace{1.5cm}
-T_X^{IJ}\bar{\eta}\tilde{\nu}_R^{*I}\tilde{\nu}_R^{*J}
+\epsilon_{ij}T^{IJ}_{\nu}H_u^i\tilde{\nu}_R^{I*}\tilde{l}_j^J
-m_{\eta}^2|\eta|^2-m_{\bar{\eta}}^2|\bar{\eta}|^2-m_S^2S^2\nonumber\\&&\hspace{1.5cm}
-(m_{\tilde{\nu}_R}^2)^{IJ}\tilde{\nu}_R^{I*}\tilde{\nu}_R^{J}
-\frac{1}{2}\Big(M_X\lambda^2_{\tilde{X}}+2M_{BB^\prime}\lambda_{\tilde{B}}\lambda_{\tilde{X}}\Big)+h.c~~.
\end{eqnarray}
$\mathcal{L}_{soft}^{MSSM}$ represent the soft breaking terms of MSSM.
In the $U(1)_X$SSM, a new effect never seen in the MSSM occurs: the gauge kinetic mixing produced by the two Abelian groups $U(1)_Y$ and
$U(1)_X$. In general, the covariant derivatives of $U(1)_X$SSM can be written as \cite{UX4,BL1,BL2,Gmass}
\begin{eqnarray}
%%%%%%%%%%%%%%%%%%%%%%%%%%%%%%%%%%%%%%%%%%%%%%%%%%%%%%%%%%%%%%%%%%%%
&&D_\mu=\partial_\mu-i\left(\begin{array}{cc}Y,&X\end{array}\right)
\left(\begin{array}{cc}g_{Y},&g{'}_{{YX}}\\g{'}_{{XY}},&g{'}_{{X}}\end{array}\right)
\left(\begin{array}{c}A_{\mu}^{\prime Y} \\ A_{\mu}^{\prime X}\end{array}\right)\;.
%%%%%%%%%%%%%%%%%%%%%%%%%%%%%%%%%%%%%%%%%%%%%%%%%%%%%%%%%%%%%%%%%%%%
\label{gauge1}
\end{eqnarray}
Under the condition that the two Abelian gauge groups are not broken, we can change the basis of the above equation by rotation matrix $R$,
with $A_{\mu}^{\prime Y}$ and $A^{\prime X}_\mu$ representing the gauge fields of $U(1)_Y$ and $U(1)_X$ respectively,
\begin{eqnarray}
&&D_\mu=\partial_\mu-i\left(\begin{array}{cc}Y^Y,&Y^X\end{array}\right)
\left(\begin{array}{cc}g_{Y},&g{'}_{{YX}}\\g{'}_{{XY}},&g{'}_{{X}}\end{array}\right)R^TR
\left(\begin{array}{c}A_{\mu}^{\prime Y} \\ A_{\mu}^{\prime X}\end{array}\right)\;.
\end{eqnarray}
Then combined with our redefined
\begin{eqnarray}
%%%%%%%%%%%%%%%%%%%%%%%%%%%%%%%%%%%%%%%%%%%%%%%%%%%%%%%%%%%%%%%%%%%%
&&\left(\begin{array}{cc}g_{Y},&g{'}_{{YX}}\\g{'}_{{XY}},&g{'}_{{X}}\end{array}\right)
R^T=\left(\begin{array}{cc}g_{1},&g_{{YX}}\\0,&g_{{X}}\end{array}\right)~\text{and}~~
%%%%%%%%%%%%%%%%%%%%%%%%%%%%%%%%%%%%%%%%%%%%%%%%%%%%%%%%%%%%%%%%%%%%
%%%%%%%%%%%%%%%%%%%%%%%%%%%%%%%%%%%%%%%%%%%%%%%%%%%%%%%%%%%%%%%%%%%%
R\left(\begin{array}{c}A_{\mu}^{\prime Y} \\ A_{\mu}^{\prime X}\end{array}\right)
=\left(\begin{array}{c}A_{\mu}^{Y} \\ A_{\mu}^{X}\end{array}\right)\;,
\end{eqnarray}
 we get the covariant derivatives of the $U(1)_X$SSM that changes the base:
 \begin{eqnarray}
%%%%%%%%%%%%%%%%%%%%%%%%%%%%%%%%%%%%%%%%%%%%%%%%%%%%%%%%%%%%%%%%%%%%
&&D_\mu=\partial_\mu-i\left(\begin{array}{cc}Y^Y,&Y^X\end{array}\right)
\left(\begin{array}{cc}g_{1},&g_{{YX}}\\0,&g_{{X}}\end{array}\right)
\left(\begin{array}{c}A_{\mu}^{Y} \\ A_{\mu}^{X}\end{array}\right)\;.
%%%%%%%%%%%%%%%%%%%%%%%%%%%%%%%%%%%%%%%%%%%%%%%%%%%%%%%%%%%%%%%%%%%%
\end{eqnarray}

Three neutral gauge bosons $A^{X}_\mu,~A^Y_\mu$ and $V^3_\mu$ mix together at the tree level, whose mass matrix
is shown in the basis $(A^Y_\mu, V^3_\mu, A^{X}_\mu)$
\begin{eqnarray}
&&\left(\begin{array}{*{20}{c}}
\frac{1}{8}g_{1}^2 v^2 &~~~ -\frac{1}{8}g_{1}g_{2} v^2 & ~~~\frac{1}{8}g_{1}(g_{{YX}}+g_{X}) v^2 \\
-\frac{1}{8}g_{1}g_{2} v^2 &~~~ \frac{1}{8}g_{2}^2 v^2 & ~~~~-\frac{1}{8}g_{2}g_{{YX}} v^2\\
\frac{1}{8}g_{1}(g_{{YX}}+g_{X}) v^2 &~~~ -\frac{1}{8}g_{2}(g_{{YX}}+g_{X}) v^2 &~~~~ \frac{1}{8}(g_{{YX}}+g_{X})^2 v^2+\frac{1}{8}g_{{X}}^2
\xi^2
\end{array}\right),\label{gauge matrix}
\end{eqnarray}
with $v^2=v_{u}^2+v_{d}^2$ and $\xi^2=v_\eta^2+v_{\bar{\eta}}^2$.

To get mass eigenvalues  of the matrix in Eq.(\ref{gauge matrix}), we use
 two mixing angles $\theta_{W}$ and $\theta_{W}'$.
 $\theta_{W}$ is the Weinberg angle and the new mixing angle $\theta_{W}'$ is
 defined by the following formula
\begin{eqnarray}
\sin^2\theta_{W}'\!=\!\frac{1}{2}\!-\!\frac{[(g_{{YX}}+g_{X})^2-g_{1}^2-g_{2}^2]v^2+
4g_{X}^2\xi^2}{2\sqrt{[(g_{{YX}}+g_{X})^2+g_{1}^2+g_{2}^2]^2v^4\!+\!8g_{X}^2[(g_{{YX}}+g_{X})^2\!-\!g_{1}^2\!-\!g_{2}^2]v^2\xi^2\!+\!16g_{X}^4\xi^4}}.
\end{eqnarray}
$\theta_{W}'$ appears in the couplings involving $Z$ and $Z^{\prime}$.
The exact eigenvalues of Eq.(\ref{gauge matrix}) are deduced
\begin{eqnarray}
&&m_\gamma^2=0,\nonumber\\
&&m_{Z,{Z^{'}}}^2=\frac{1}{8}\Big([g_{1}^2+g_2^2+(g_{{YX}}+g_{X})^2]v^2+4g_{X}^2\xi^2 \nonumber\\
&&\hspace{1.1cm}\mp\sqrt{[g_{1}^2+g_{2}^2+(g_{{YX}}+g_{X})^2]^2v^4\!+\!8[(g_{{YX}}+g_{X})^2\!-\!g_{1}^2\!-\!
g_{2}^2]g_{X}^2v^2\xi^2\!+\!16g_{X}^4\xi^4}\Big).
\end{eqnarray}
The other used mass matrices can be found in the works \cite{UU1,20}.

In addition, there are some couplings that need to be used later:
\begin{eqnarray}
&&\mathcal{L}_{\bar{d}g\tilde{d}}=\bar{d_{i}}\sqrt{2} g_3 \Big(\tilde{D}_i^R P_L - \tilde{D}_i^L P_R\Big)\Lambda_G,
\\&&
\mathcal{L}_{\bar{u}dH}=\bar{u}_i
\Big(Y_{u,i}H^+_uP_L+Y_{d,j}H^+_dP_R\Big)V_{ij}d_{j},
\\&&
\mathcal{L}_{\bar{\chi}^- d\tilde{u}}=\Big[(\tilde{H}_u^+Y_{u,i}\tilde{U}_i^{R*} - g_2\tilde{W}^+\tilde{U}_i^{L*} )P_L+Y_{d,j} \tilde{U}_i^{L*}
\tilde{H}_d^+P_R \Big]V_{ij}d_{j},
\\&&
\mathcal{L}_{\bar{d}\chi^0\tilde{D}}=\bar{d}_{i}
\Big\{-\frac{1}{6} [2\sqrt{2} g_1 \tilde{B}\tilde{D}_{i}^R + \sqrt{2}(2g_{YX}+3g_X)\lambda_{\tilde{X}}\tilde{D}_i^R + 6Y_{d,i}
\tilde{H}_d^0\tilde{D}_i^L]P_L
\nonumber\\&&\hspace{1.5cm}- \frac{1}{6} [6 Y_{d,i}\tilde{D}_i^R\tilde{H}_d^0 + \sqrt{2}\tilde{D}_i^L(g_1 \tilde{B} + g_{YX}
\lambda_{\tilde{X}} - 3 g_2\tilde{W}^0 ) ]P_R \Big\}.
\end{eqnarray}

\section{Formulation}
At scale $\mu=O(m_b)$, the effective Hamiltonian of the flavor transition process $b\rightarrow s\gamma$ has the following form \cite{ham}:
\begin{eqnarray}
\mathcal {H}_{eff}(b\to s\gamma)=-\frac{4G_F}{\sqrt{2}}V_{32}^*V_{33}
\Big[C_1Q_1^c+C_2Q_2^c+\sum_{i=3}^6C_iQ_i+\sum_{i=7}^8(C_iQ_i+\tilde{C}_i\tilde{Q}_i)\Big],
\label{eff-hamiltonian}
\end{eqnarray}
and the operators are given by Refs. \cite{oper1,oper2,oper3}:
\begin{eqnarray}
%%%%%%%%%%%%%%%%%%%%%%%%%%%%%%%%%%%%%%%%%%%%%%%%%%%%%%%%%%%%%%%%%%%%%%
\mathcal{O}_1^c
	&=&(\bar{s}_L\gamma_\mu T^ab_L)(\bar{c}_L\gamma^\mu T^ab_L),
~~~~~\mathcal{O}_2^c=(\bar{s}_L\gamma_\mu b_L)(\bar{c}_L\gamma^\mu b_L),\nonumber\\
\mathcal{O}_3
	&=&(\bar{s}_L\gamma_\mu b_L)\sum_q(\bar{q}\gamma^\mu q),
~~~~~~~~~~\mathcal{O}_4=(\bar{s}_L\gamma_\mu T^ab_L)\sum_q(\bar{q}\gamma^\mu T^a q),\nonumber\\
\mathcal{O}_5
	&=&(\bar{s}_L\gamma_\mu \gamma_\nu\gamma_\rho b_L)\sum_q(\bar{q}\gamma^\mu \gamma^\nu\gamma^\rho q),\nonumber\\
\mathcal{O}_6
	&=&(\bar{s}_L\gamma_\mu \gamma_\nu\gamma_\rho T^a b_L)\sum_q(\bar{q}\gamma^\mu \gamma^\nu\gamma^\rho T^a q),\nonumber\\
\mathcal{O}_7
	&=&e/g_3^2m_b(\bar{s}_L\sigma_{\mu\nu}b_R)F^{\mu\nu},
~~~~~~~\mathcal{O}_8=1/g_3^2m_b(\bar{s}_L\sigma_{\mu\nu}T^ab_R)G^{a,\mu\nu},\nonumber\\
\tilde{\mathcal{O}}_7
	&=&e/g_3^2m_b(\bar{s}_R\sigma_{\mu\nu}b_L)F^{\mu\nu},
~~~~~~~\tilde{\mathcal{O}}_8=1/g_3^2m_b(\bar{s}_R\sigma_{\mu\nu}T^ab_L)G^{a,\mu\nu}.
\end{eqnarray}
Through the amplitudes of the Feynman diagrams involved in the process $b\rightarrow s\gamma$, we can extract the coefficients of these
operators. Actually, when we calculate the branching ratio with formula presented in Ref. \cite{ham}:
\begin{eqnarray}
%%%%%%%%%%%%%%%%%%%%%%%%%%%%%%%%%%%%%%%%%%%%%%%%%%%%%%%%%%%%%%%%%%%%%%
&& BR(\bar{B}\rightarrow X_s\gamma)_{NP}\nonumber\\
&&=10^{-4}\times\left\{(3.36\pm0.23)+\frac{16\pi^2a_{77}}{\alpha_s^2(\mu_b)}\big[|C_{7,NP}(\mu_{EW})|^2+|\tilde{C}_{7,NP}(\mu_{EW})|^2\big]\right.\nonumber\\
&&+\frac{16\pi^2a_{88}}{\alpha_s^2(\mu_b)}\big[|C_{8,NP}(\mu_{EW})|^2+|\tilde{C}_{8,NP}(\mu_{EW})|^2\big]\nonumber\\
&&+\frac{ 4\pi}{\alpha_s(\mu_b)}\mbox{Re}\big[a_7C_{7,NP}(\mu_{EW})+a_8C_{8,NP}(\mu_{EW})\big.\nonumber\\
&&+\left.\big.\frac{ 4\pi
a_{78}}{\alpha_s(\mu_b)}\big(C_{7,NP}(\mu_{EW})C_{8,NP}(\mu_{EW})+\tilde{C}_{7,NP}(\mu_{EW})\tilde{C}_{8,NP}(\mu_{EW})\big)\big]\right\},
%%%%%%%%%%%%%%%%%%%%%%%%%%%%%%%%%%%%%%%%%%%%%%%%%%%%%%%%%%%%%%%%%%%%%%
\label{BRNP}
\end{eqnarray}
only the coefficients of $\mathcal{O}_{7,8}$ and $\tilde{\mathcal{O}}_{7,8}$ are needed. In Eq.($\ref{BRNP}$), the first term is the SM
prediction, and $C_{7,NP}(\mu_{EW})$, $C_{8,NP}(\mu_{EW})$, $\tilde{C}_{7,NP}(\mu_{EW})$ and $\tilde{C}_{8,NP}(\mu_{EW})$ indicate Wilson
coefficients at electroweak scale. In this representation, the effect evolved to the hadron scale $\mu\sim m_b$ has been
included in the coefficients $a_{7,8,77,88,78}$. The numerical values of these coefficients are shown in Table~\ref{coeff-a}.

\begin{table}[tbh]
\caption{Numerical values for the coefficients $a_{7,8,77,88,78}$ at electroweak scale.}
\begin{tabular}{@{}c|c|c|c|c@{}} \toprule
$a_7$ & $a_8$ & $a_{77}$ & $a_{88}$ & $a_{78}$\\
\colrule
$-7.184+0.612i$ & $-2.225-0.557i$ & $4.743$ & $0.789$ & $2.454-0.884i$\\
\botrule
\end{tabular}
\label{coeff-a}
\end{table}

\subsection{Using MIA to calculate $b\rightarrow s\gamma$ in $U(1)_X$SSM}
\begin{figure}[h]
\centering
\subfigure[]{
\setlength{\unitlength}{5.0mm}
\includegraphics[width=1.9in]{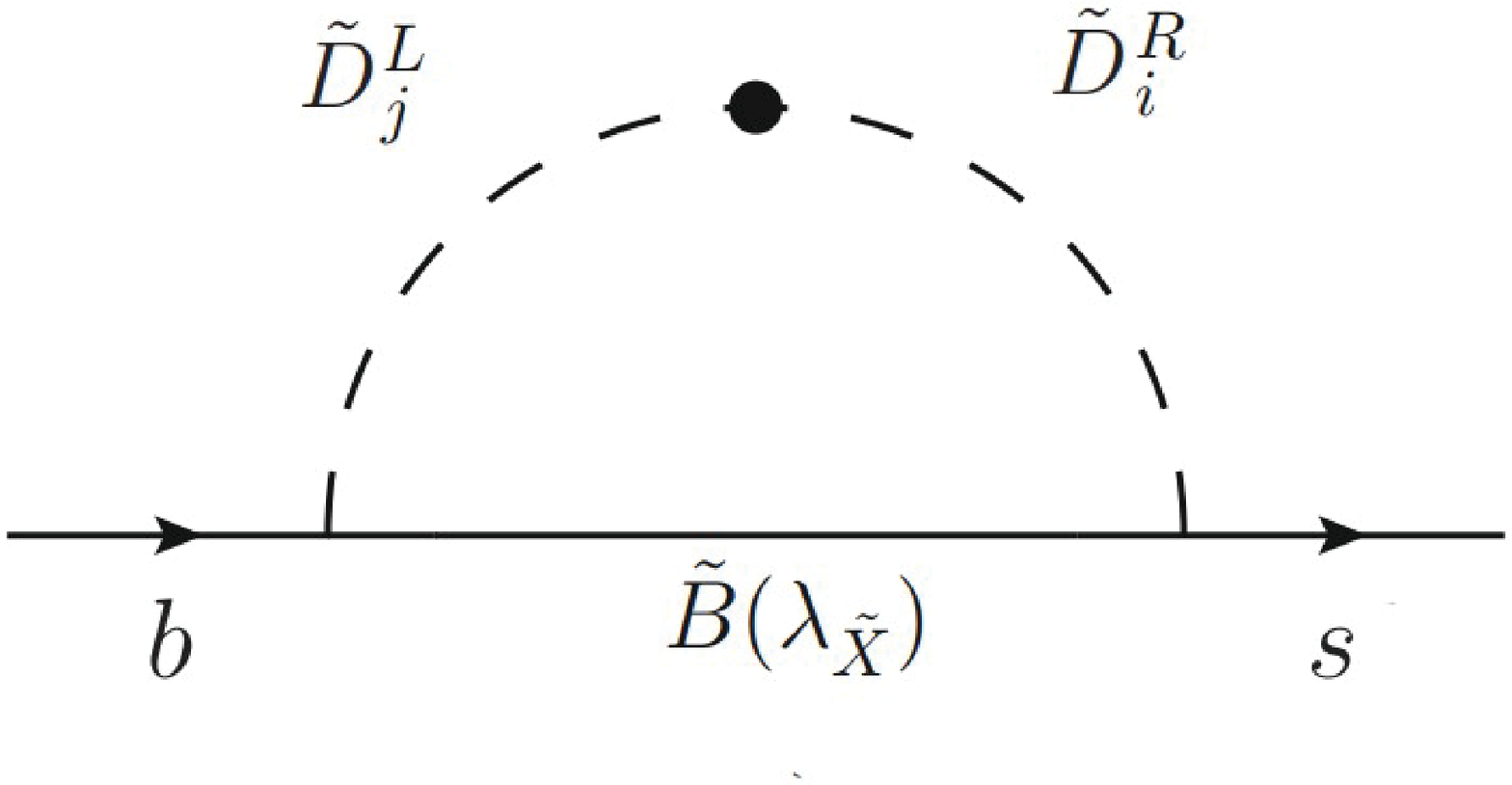}
\label{Fig1}
}
\subfigure[]{
\setlength{\unitlength}{5.0mm}
\includegraphics[width=1.9in]{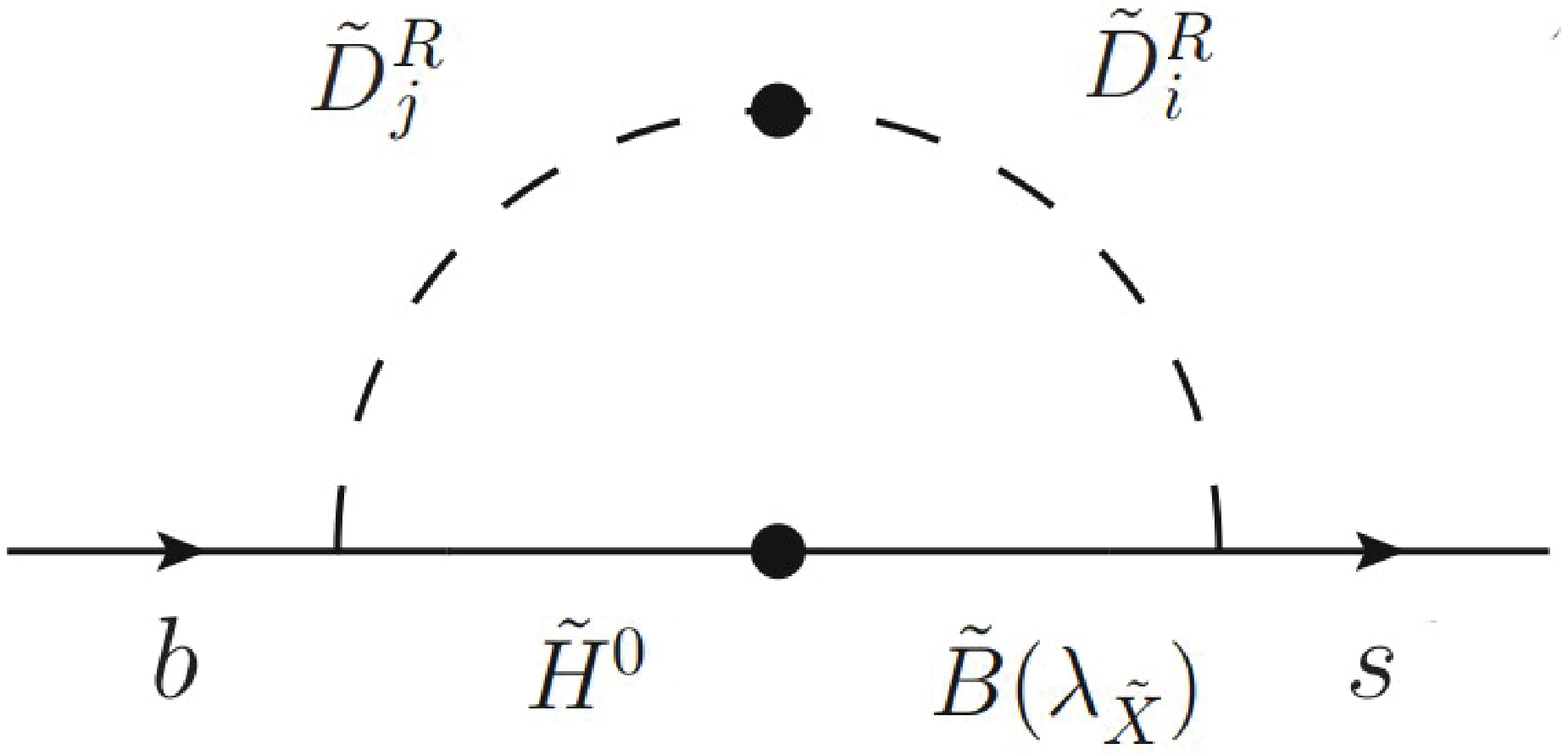}
\label{Fig2}
}
\subfigure[]{
\setlength{\unitlength}{5.0mm}
\includegraphics[width=1.9in]{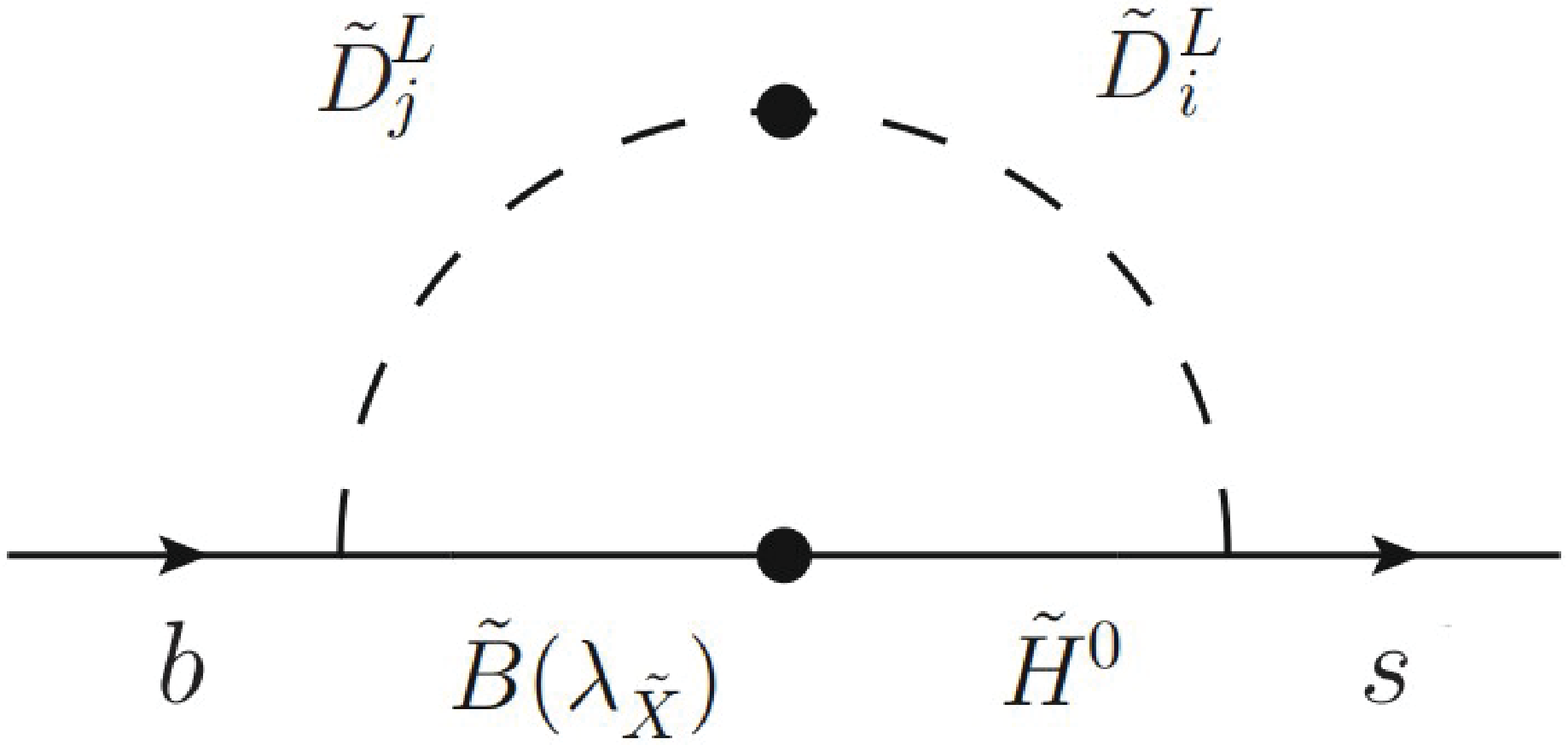}
\label{Fig3}
}
\subfigure[]{
\setlength{\unitlength}{5.0mm}
\includegraphics[width=1.9in]{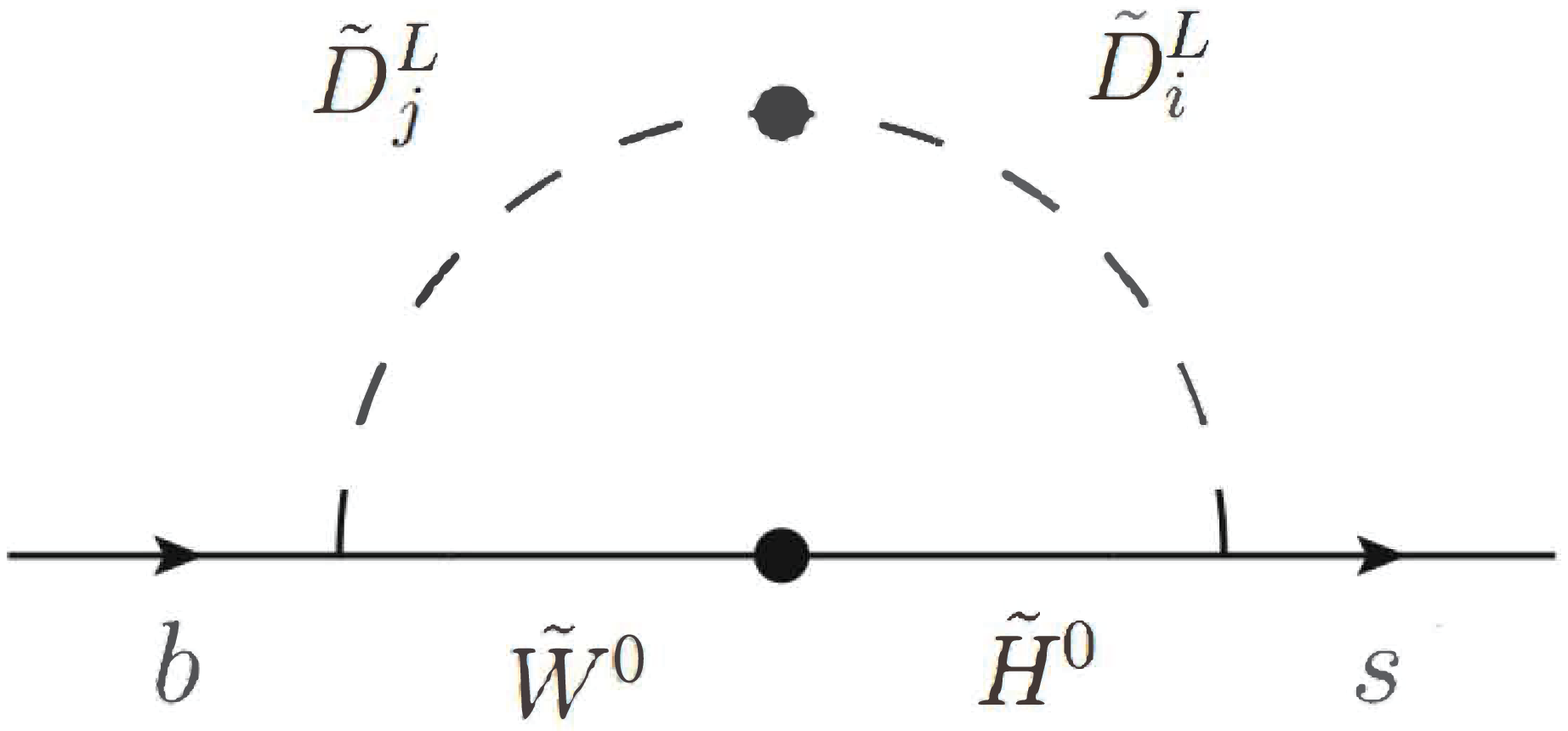}
\label{Fig4}
}
\subfigure[]{
\setlength{\unitlength}{5.0mm}
\includegraphics[width=1.9in]{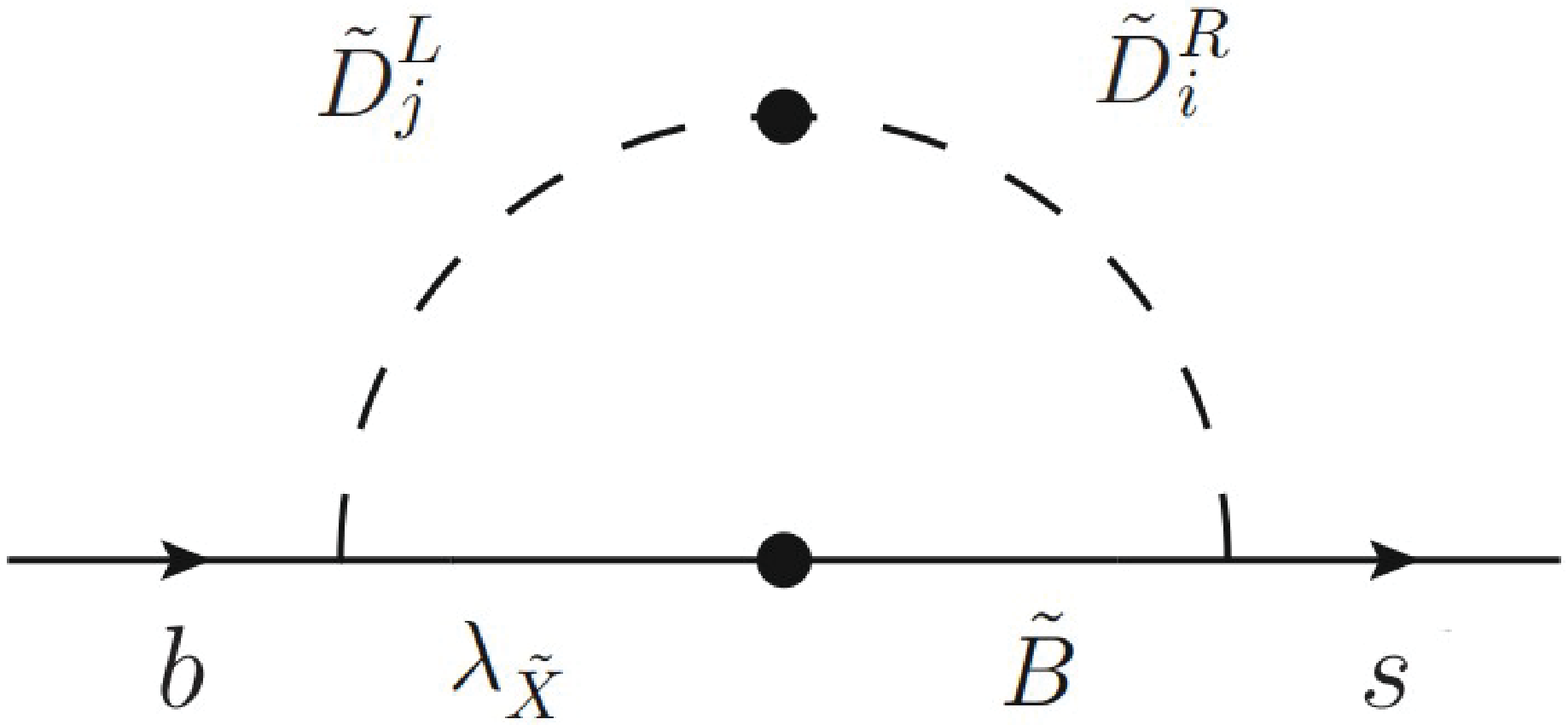}
\label{Fig5}
}
\subfigure[]{
\setlength{\unitlength}{5.0mm}
\includegraphics[width=1.9in]{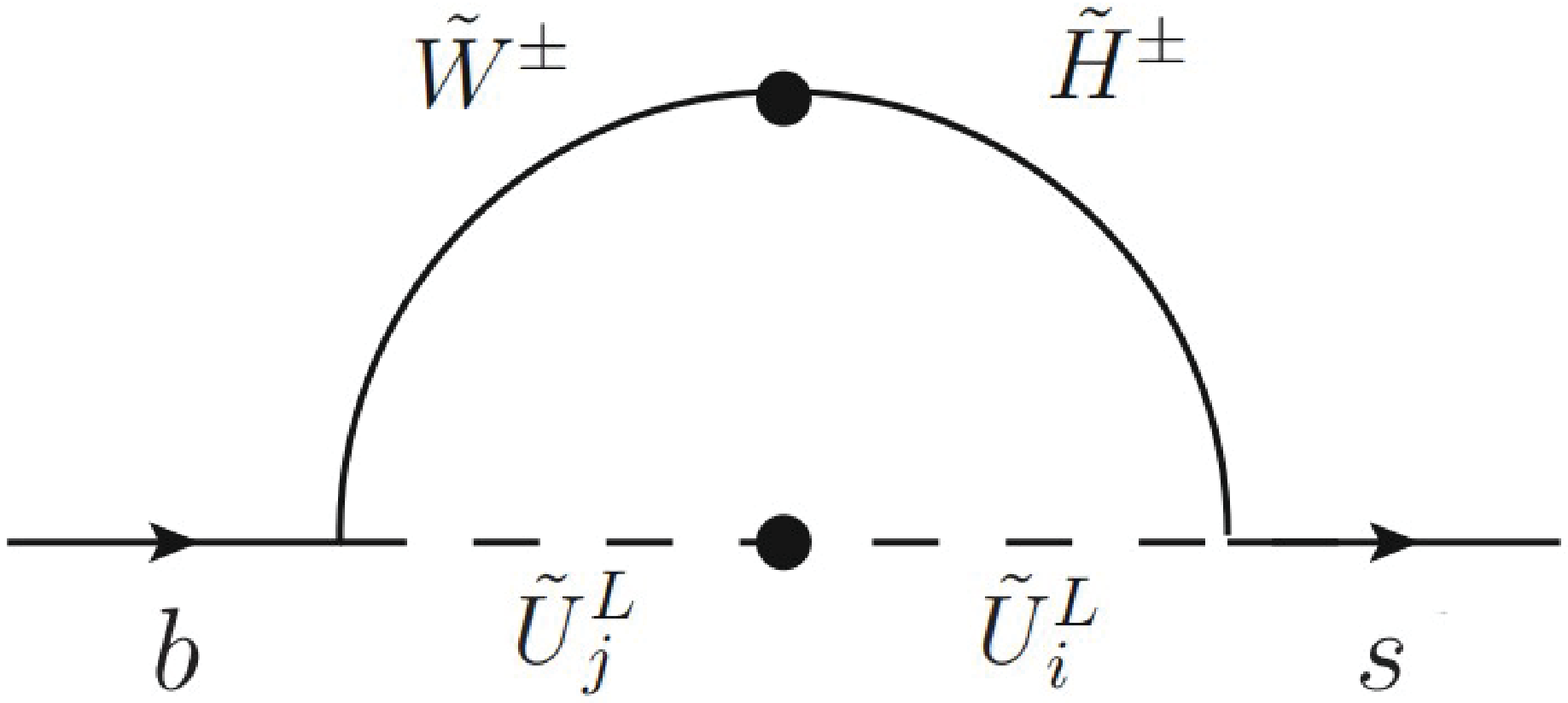}
\label{Fig6}
}
\subfigure[]{
\setlength{\unitlength}{5.0mm}
\includegraphics[width=1.9in]{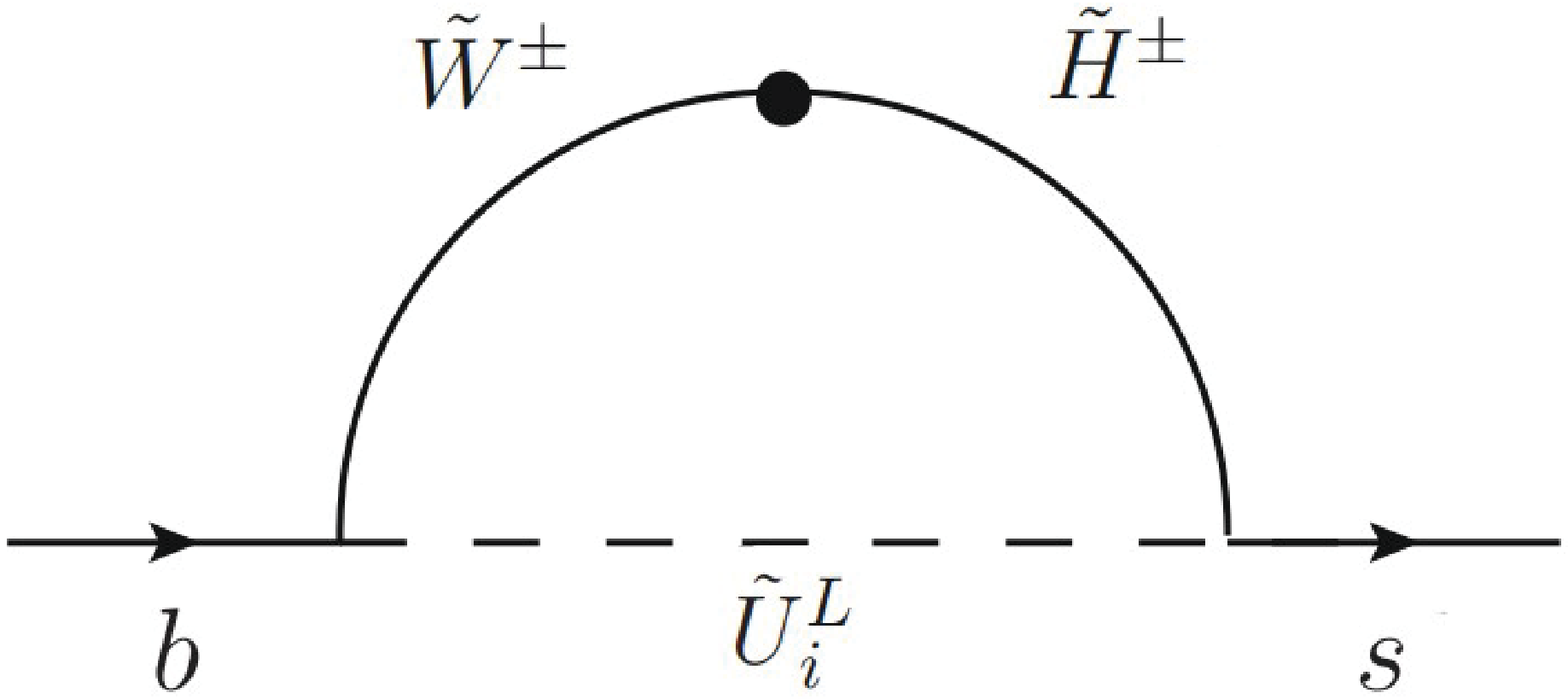}
\label{Fig7}
}
\subfigure[]{
\setlength{\unitlength}{5.0mm}
\includegraphics[width=1.9in]{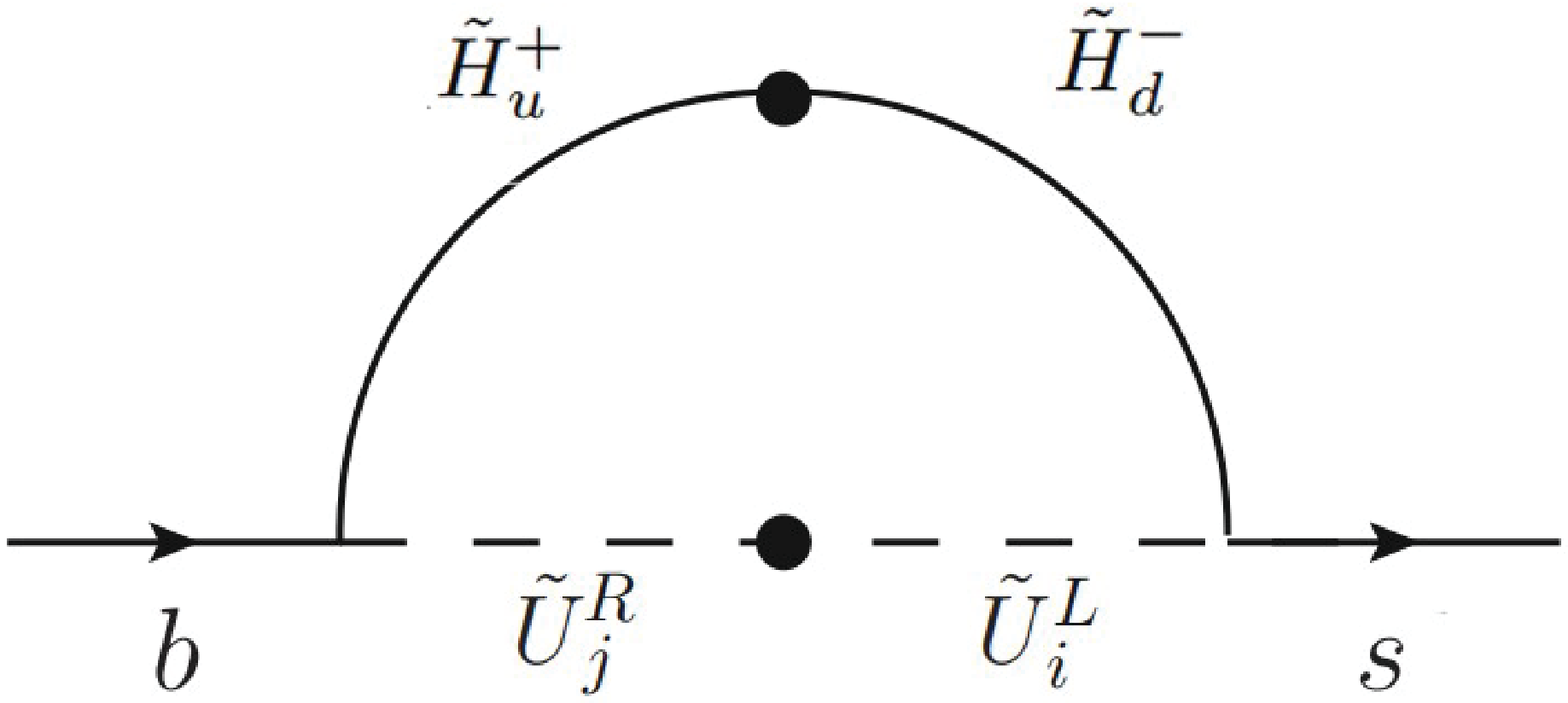}
\label{Fig8}
}
\subfigure[]{
\setlength{\unitlength}{5.0mm}
\includegraphics[width=1.9in]{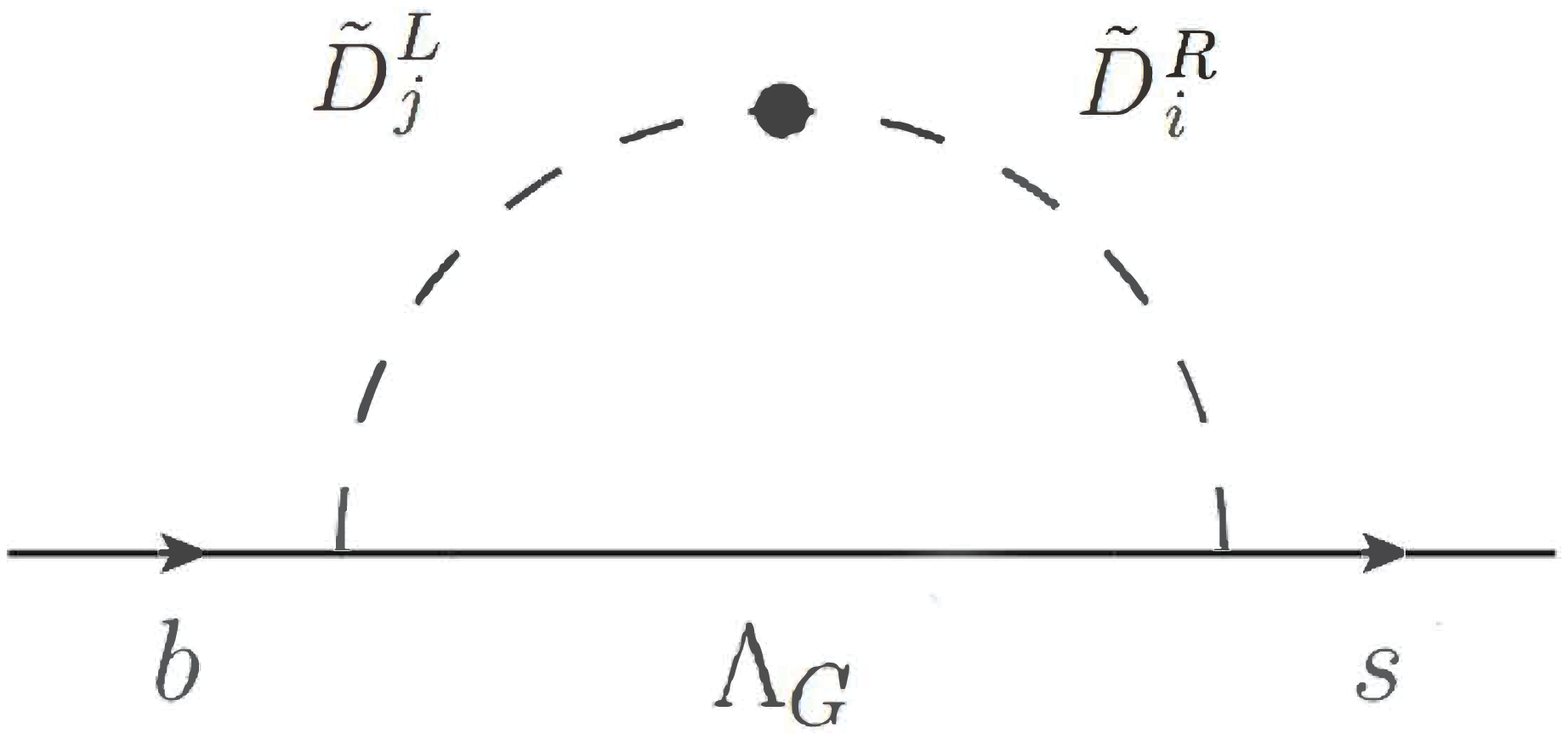}
\label{Fig9}
}
\caption{Self-energy Feynman diagrams for $b\rightarrow s\gamma$ in the MIA.}\label{N1}
\end{figure}

The self-energy Feynman diagrams under the $U(1)_X$SSM are obtained by MIA in Fig.~\ref{N1}.
To obtain the branching ratio, we research one-loop diagrams in Fig.~\ref{N1}. To obtain triangle diagrams, the photon should be attached to all
inner lines with electric charge. In this case, the coefficients of $\mathcal{O}_7$ and $\tilde{\mathcal{O}}_7$ can be extracted from the
amplitudes of these diagrams of $b\to s\gamma$. Similarly, the coefficients of $\mathcal{O}_8$ and $\tilde{\mathcal{O}}_8$ can be obtained from
the diagrams of $b\rightarrow sg$ with gluons attached to all the inner lines with color charge.
We give the following two basic functions:
\begin{eqnarray}
&&f(x,y,z)=-\frac{\rm i}{16 \pi ^2} \big[\frac{x \log x}{(x-y) (x-z)}-\frac{y \log y}{(y-x) (y-z)}-\frac{z \log z}{(z-x)
   (z-y)}\big],\label{f1}
%%%%%%%%%%%%%%%%%%%%%%%%%%%%%%%%%%%%%%%%%%%%%%%%%%
\\&&g(x,y,z,t)=\frac{1}{z-t}\Big(f(x,y,z)-f(x,y,t)\Big).\label{g1}
\end{eqnarray}

1. The one-loop contributions from $\tilde{B}(\lambda_{\tilde{X}})$-$\tilde{D}^L_j$-$\tilde{D}^R_i$ $(i=2,~j=3)$.
\begin{eqnarray}
&&C_7^1(\tilde{D}^L_j,\tilde{D}^R_i,\tilde{B})=\frac{1}{216}\frac{{T_d}^{ij}v\cos{\beta}}{\mathcal{A}m_b\Lambda^3}g_3^2g_1^2
\sqrt{x_1}I_1(x_1,x_{\tilde{D}^L_j},x_{\tilde{D}^R_i}),\label{MIAB}
%%%%%%%%%%%%%%%%%%%%%%%%%%%%%%%%%%%%%%%%%%%%%%%%%%
\\&&C_7^2(\tilde{D}^L_j,\tilde{D}^R_i, \lambda_{\tilde{X}})
=\frac{1}{432}\frac{{T_d}^{ij}v\cos{\beta}}{\mathcal{A}m_b\Lambda^3}g_3^2(2g_{YX}^2+3g_{YX}g_X)
\sqrt{x_{\lambda_{\tilde{X}}}}I_1(x_{\lambda_{\tilde{X}}},x_{\tilde{D}^L_j},x_{\tilde{D}^R_i}),\label{MIAX}
%%%%%%%%%%%%%%%%%%%%%%%%%%%%%%%%%%%%%%%%%%%%%%%%%%
\\&&C_8^1(\tilde{D}^L_j,\tilde{D}^R_i,\tilde{B})
=\frac{1}{216}\frac{{T_d}^{ij}v\cos{\beta}}{\mathcal{A}m_b\Lambda^3}g_3^3g_1^2
\sqrt{x_1}I_1(x_1,x_{\tilde{D}^L_j},x_{\tilde{D}^R_i}),\label{MIAB2}
%%%%%%%%%%%%%%%%%%%%%%%%%%%%%%%%%%%%%%%%%%%%%%%%%%
\\&&C_8^2(\tilde{D}^L_j,\tilde{D}^R_i, \lambda_{\tilde{X}})
=\frac{1}{432}\frac{{T_d}^{ij}v\cos{\beta}}{\mathcal{A}m_b\Lambda^3}g_3^3(2g_{YX}^2+3g_{YX}g_X)
\sqrt{x_{\lambda_{\tilde{X}}}}I_1(x_{\lambda_{\tilde{X}}},x_{\tilde{D}^L_j},x_{\tilde{D}^R_i}),\label{MIAX2}
\end{eqnarray}
in which, $m$ stands for the particle mass, with $x=\frac{m^2}{\Lambda^2}$. The coefficient $\mathcal{A}$ is equal to
$\frac{1}{G_FV_{32}^*V_{33}}$ and $\Lambda$ represents the energy scale. The function $I_1(x,y,z)$ is
\begin{eqnarray}
&&I_1(x,y,z)=-\Big(\frac{z}{2}\frac{\partial^2}{\partial z^2}+\frac{y}{2}\frac{\partial^2}{\partial y^2}+\frac{\partial}{\partial
z}+y\frac{\partial^2}{\partial y \partial z}\Big)f(x,y,z).
\end{eqnarray}

2. The one-loop contributions from $\tilde{B}(\lambda_{\tilde{X}})$-$\tilde{H}^0$-$\tilde{D}^R_j$-$\tilde{D}^R_i$ $(i=2,~j=3)$.

Fig.s~\ref{Fig2}-\ref{Fig5} are all computed with the same method of coefficients and the functions required in the process, so here we take
Fig.~\ref{Fig2} as an example.
\begin{eqnarray}
&&C_7^3(\tilde{D}^R_j,\tilde{D}^R_i,\tilde{B},\tilde{H}^0)
=\frac{\sqrt{2}}{72}\frac{m_s\Delta^{RR}_{ij}(\tilde{D})}{\mathcal{A}m_b\Lambda^4}g_3^2g_1^2\tan{\beta}
\sqrt{x_1x_{\mu^{\prime}_H}}I_2(x_1,x_{\tilde{D}^R_j},x_{\tilde{D}^R_i},x_{\mu^{\prime}_H}),
%%%%%%%%%%%%%%%%%%%%%%%%%%%%%%%%%%%%%%%%%%%%%%%%%%
\\&&C_7^4(\tilde{D}^R_j,\tilde{D}^R_i, \lambda_{\tilde{X}},\tilde{H}^0)
=\frac{\sqrt{2}}{288}\frac{m_s\Delta^{RR}_{ij}(\tilde{D})}{\mathcal{A}m_b\Lambda^4}g_3^2(2g_{YX}+3g_X)(g_{YX}+g_X)\nonumber\\
&&\hspace{4.3cm}\times\tan{\beta} \sqrt{2x_{\lambda_{\tilde{X}}}x_{\mu^{\prime}_H}}
I_2(x_{\lambda_{\tilde{X}}},x_{\tilde{D}^R_j},x_{\tilde{D}^R_i},x_{\mu^{\prime}_H}),
%%%%%%%%%%%%%%%%%%%%%%%%%%%%%%%%%%%%%%%%%%%%%%%%%%
\\&&C_8^3(\tilde{D}^R_j,\tilde{D}^R_i,\tilde{B},\tilde{H}^0)
=\frac{\sqrt{2}}{72}\frac{m_s\Delta^{RR}_{ij}(\tilde{D})}{\mathcal{A}m_b\Lambda^4}g_3^3g_1^2\tan{\beta}
\sqrt{x_1x_{\mu^{\prime}_H}}I_2(x_1,x_{\tilde{D}^R_j},x_{\tilde{D}^R_i},x_{\mu^{\prime}_H}),
%%%%%%%%%%%%%%%%%%%%%%%%%%%%%%%%%%%%%%%%%%%%%%%%%%
\\&&C_8^4(\tilde{D}^R_j,\tilde{D}^R_i, \lambda_{\tilde{X}},\tilde{H}^0)
=\frac{\sqrt{2}}{288}\frac{m_s\Delta^{RR}_{ij}(\tilde{D})}{\mathcal{A}m_b\Lambda^4}g_3^3(2g_{YX}+3g_X)(g_{YX}+g_X)\nonumber\\
&&\hspace{4.3cm}\times\tan{\beta}\sqrt{2x_{\lambda_{\tilde{X}}}x_{\mu^{\prime}_H}}
I_2(x_{\lambda_{\tilde{X}}},x_{\tilde{D}^R_j},x_{\tilde{D}^R_i},x_{\mu^{\prime}_H}),\label{MIAHX2}
\end{eqnarray}
in which, $\mu_{H}^\prime=\frac{\lambda_H v_S}{\sqrt{2}}+\mu$ and $x_{\mu^{\prime}_H}=\frac{\mu_{H}^{\prime2}}{\Lambda^2}$. The function
$I_2(x,y,z,t)$ is
\begin{eqnarray}
&&I_2(x,y,z,t)=-\big[\frac{z}{2}\frac{\partial^2}{\partial z^2}+\frac{y}{2}\frac{\partial^2}{\partial y^2}+\frac{\partial}{\partial
z}+y\frac{\partial^2}{\partial y \partial z}\big]g(x,y,z,t).
\end{eqnarray}

3. The one-loop contributions from $\tilde{H}^{\pm}$-$\tilde{W}^{\pm}$-$\tilde{U}^L_j$-$\tilde{U}^L_i$.
\begin{eqnarray}
&&C_7^5(\tilde{U}^L_j,\tilde{U}^L_i,\tilde{H}^{\pm},\tilde{W}^{\pm})
=\frac{\sqrt{2}}{4}\sum_{i,j=1}^{3}\frac{m_s\Delta^{LL}_{ij}(\tilde{U})}{\mathcal{A}vm_b\Lambda^3}V^*_{2j}V_{3i}g_3^2g_2
\Big\{\sqrt{x_2}I_3(x_{\mu^{\prime}_H},x_{\tilde{U}^L_j},x_{\tilde{U}^L_i},x_{2})\nonumber\\
&&\hspace{4.3cm}+x_2\sqrt{x_{\mu^{\prime}_H}}\tan{\beta}I_4(x_{\mu^{\prime}_H},x_{\tilde{U}^R_j},x_{\tilde{U}^R_i},x_{2})\Big\},\label{MIAHW}
\end{eqnarray}
\begin{eqnarray}
&&C_7^6(\tilde{U}^L_j,\tilde{U}^L_i,\tilde{H}^{\pm},\tilde{W}^{\pm})
=\frac{\sqrt{2}}{6}\sum_{i,j=1}^{3}\frac{m_s\Delta^{LL}_{ij}(\tilde{U})}{\mathcal{A}vm_b\Lambda^3}V^*_{2j}V_{3i}g_3^2g_2
\Big\{\sqrt{x_2}I_3(x_{\mu^{\prime}_H},x_{\tilde{U}^L_j},x_{\tilde{U}^L_i},x_{2})\nonumber\\
&&\hspace{4.3cm}+x_2\sqrt{x_{\mu^{\prime}_H}}\tan{\beta}I_2(x_{\mu^{\prime}_H},x_{\tilde{U}^L_j},x_{\tilde{U}^L_i},x_{2})\Big\},\label{MIAWH}
%%%%%%%%%%%%%%%%%%%%%%%%%%%%%%%%%%%%%%%%%%%%%%%%%%
\\&&C_8^5(\tilde{U}^L_j,\tilde{U}^L_i,\tilde{H}^{\pm},\tilde{W}^{\pm})
=\frac{\sqrt{2}}{6}\sum_{i,j=1}^{3}\frac{m_s\Delta^{LL}_{ij}(\tilde{U})}{\mathcal{A}vm_b\Lambda^3}V^*_{2j}V_{3i}g_3^3g_2
\Big\{\sqrt{x_2}I_3(x_{\mu^{\prime}_H},x_{\tilde{U}^L_j},x_{\tilde{U}^L_i},x_{2})\nonumber\\
&&\hspace{4.3cm}+x_2\sqrt{x_{\mu^{\prime}_H}}\tan{\beta}I_2(x_{\mu^{\prime}_H},x_{\tilde{U}^L_j},x_{\tilde{U}^L_i},x_{2})\Big\}.\label{MIAWH2}
\end{eqnarray}
The functions $I_3(x,y,z,t)$ and $I_4(x,y,z,t)$ are
\begin{eqnarray}
&&I_3(x,y,z,t)=-\big[1+2y\frac{\partial}{\partial y}+2z\frac{\partial}{\partial z}+\frac{y^2}{2}\frac{\partial^2}{\partial
y^2}+\frac{z^2}{2}\frac{\partial^2}{\partial z^2}+yz\frac{\partial z^2}{\partial y \partial z}\big]g(x,y,z,t),
\\&&I_4(x,y,z,t)=\big[\frac{\partial}{\partial y}+2\frac{\partial}{\partial z}+\frac{y}{2}\frac{\partial^2}{\partial
y^2}+\frac{z}{2}\frac{\partial^2}{\partial z^2}+y\frac{\partial^2}{\partial y \partial z}\big]g(x,y,z,t).
\end{eqnarray}
Similar to the third case, $C_7^7$, $C_7^8$, $C_8^6$ can be extracted from Fig.~\ref{Fig8}, and calculation process and form are the same as
those of $C_7^5$, $C_7^6$, $C_8^5$. We don't repeat them here.

4. The one-loop contributions from $\tilde{H}^{\pm}$-$\tilde{W}^{\pm}$-$\tilde{U}^L_i$.
\begin{eqnarray}
&&C_7^7(\tilde{U}^L_i,\tilde{H}^{\pm},\tilde{W}^{\pm})
=\frac{\sqrt{2}}{4}\sum_{i=1}^{3}\frac{m_s}{\mathcal{A}vm_b\Lambda}V^*_{2i}V_{3i}g_3^2g_2
\Big\{x_2\sqrt{x_{\mu^{\prime}_H}}\tan{\beta}I_5(x_{\tilde{U}^L_i},x_{2},x_{\mu^{\prime}_H})\nonumber\\
&&\hspace{3.6cm}+\sqrt{x_2}I_6(x_{\tilde{U}^L_i},x_{2},x_{\mu^{\prime}_H})\Big\},\label{MIAHB}
%%%%%%%%%%%%%%%%%%%%%%%%%%%%%%%%%%%%%%%%%%%%%%%%%%
\\&&C_7^8(\tilde{U}^L_i,\tilde{H}^{\pm},\tilde{W}^{\pm})=\frac{\sqrt{2}}{6}\sum_{i=1}^{3}
\frac{m_s}{\mathcal{A}vm_b\Lambda}V^*_{2i}V_{3i}g_3^2g_2\Big\{x_2\sqrt{x_{\mu^{\prime}_H}}\tan{\beta}
I_{7}(x_{\mu^{\prime}_H},x_{2},x_{\tilde{U}^L_i})\nonumber\\
&&\hspace{3.6cm}-\sqrt{x_2}I_{8}(x_{\mu^{\prime}_H},x_{2},x_{\tilde{U}^L_i})\Big\},\label{MIAHB}
%%%%%%%%%%%%%%%%%%%%%%%%%%%%%%%%%%%%%%%%%%%%%%%%%%
\\&&C_8^6(\tilde{U}^L_i,\tilde{H}^{\pm},\tilde{W}^{\pm})=\frac{\sqrt{2}}{6}\sum_{i=1}^{3}
\frac{m_s}{\mathcal{A}vm_b\Lambda}V^*_{2i}V_{3i}g_3^3g_2\Big\{x_2\sqrt{x_{\mu^{\prime}_H}}\tan{\beta}
I_{7}(x_{\mu^{\prime}_H},x_{2},x_{\tilde{U}^L_i})\nonumber\\
&&\hspace{3.6cm}-\sqrt{x_2}I_{8}(x_{\mu^{\prime}_H},x_{2},x_{\tilde{U}^L_i})\Big\}.\label{MIAHB2}
\end{eqnarray}
The functions $I_5(x,y,z)$, $I_6(x,y,z)$, $I_{7}(x,y,z)$ and $I_{8}(x,y,z)$ are
\begin{eqnarray}
&&I_5(x,y,z)=\big[\frac{\partial}{\partial y}+2\frac{\partial}{\partial z}+\frac{y}{2}\frac{\partial^2}{\partial
y^2}+\frac{z}{2}\frac{\partial^2}{\partial z^2}+y\frac{\partial^2}{\partial y \partial z}\big]f(x,y,z),
%%%%%%%%%%%%%%%%%%%%%%%%%%%%%%%%%%%%%%%%%%%%
\\&&I_6(x,y,z)=\big[1+2y\frac{\partial}{\partial y}+2z\frac{\partial}{\partial z}+\frac{y^2}{2}\frac{\partial^2}{\partial
y^2}+\frac{z^2}{2}\frac{\partial^2}{\partial z^2}+yz\frac{\partial z^2}{\partial y \partial z}\big]f(x,y,z),
%%%%%%%%%%%%%%%%%%%%%%%%%%%%%%%%%%%%%%%%%%%%
\\&&I_{7}(x,y,z)=\big[\frac{\partial}{\partial z}+\frac{z}{2}\frac{\partial^2}{\partial
z^2}\big]f(x,y,z),~~~~I_{8}(x,y,z)=\big[1+z\frac{\partial}{\partial z}\big]f(x,y,z).
\end{eqnarray}

5. The one-loop contributions from $~\Lambda_G$-$\tilde{D}^L_j$-$\tilde{D}^R_i$ $(i=2,~j=3)$.
\begin{eqnarray}
&&C_7^9(\tilde{D}^L_j,\tilde{D}^R_i,\Lambda_G)=
-\frac{1}{48}\frac{{T_d}^{ij}v\cos{\beta}}{\mathcal{A}m_b\Lambda^3}g_3^4\sqrt{x_{\Lambda_G}}
I_1(x_{\Lambda_G},x_{\tilde{D}^L_j},x_{\tilde{D}^R_i}),\label{MIAHB}
%%%%%%%%%%%%%%%%%%%%%%%%%%%%%%%%%%%%%%%%%%%%%%%%%%
\\&&C_8^7(\tilde{D}^L_j,\tilde{D}^R_i,\Lambda_G)
=-\frac{1}{16}\frac{{T_d}^{ij}v\cos{\beta}}{\mathcal{A}m_b\Lambda^3}g_3^5\sqrt{x_{\Lambda_G}}
I_{7}(x_{\tilde{D}^L_j},x_{\tilde{D}^R_i},x_{\Lambda_G}),\label{MIAHB}
%%%%%%%%%%%%%%%%%%%%%%%%%%%%%%%%%%%%%%%%%%%%%%%%%%
\\&&C_8^8(\tilde{D}^L_j,\tilde{D}^R_i,\Lambda_G)
=-\frac{1}{48}\frac{{T_d}^{ij}v\cos{\beta}}{\mathcal{A}m_b\Lambda^3}g_3^5\sqrt{x_{\Lambda_G}}
I_1(x_{\Lambda_G},x_{\tilde{D}^L_j},x_{\tilde{D}^R_i}).\label{MIAHB2}
\end{eqnarray}
According to the above procedure, we can get the coefficients of all the diagrams in Fig.~\ref{N1}.
We add the coefficients of each diagram separately to obtain the Wilson coefficients, and import them into Eq.($\ref{BRNP}$) to obtain the branching ratio.

\subsection{Degenerate Result}
Assuming that the masses of all the superparticles are almost degenerate, we are able to more intuitively analyze the factors affecting the
flavor transition process $b\rightarrow s\gamma$. In other words, we give the one-loop results in the extreme case, where the masses for
superparticles
(~$M_1,~M_2,~\mu_H^\prime,~M_{\Lambda_G},~M_{\lambda_{\tilde{X}}},~M_{\tilde{D}_L},~M_{\tilde{D}_R},~M_{\tilde{U}_L},~M_{\tilde{U}_R},~M_{BB^\prime}
$~) are equal to $M_{SUSY}$:
\begin{eqnarray}
&&|M_1|=|M_2|=|\mu_H^\prime|=|M_{\Lambda_G}|=|M_{\lambda_{\tilde{X}}}|=|M_{BB^\prime}|
\nonumber\\&&=M_{\tilde{D}_L}=M_{\tilde{D}_R}=M_{\tilde{U}_L}=M_{\tilde{U}_R}=M_{SUSY}.
\end{eqnarray}
The functions $I_i(i=1\cdots11)$,~$\Delta^{AB}_{ij}(\tilde{Q})(A,B=R,L)$, ${T_d}^{ij}$ and ${T_u}^{ij}$ are much simplified as
\begin{eqnarray}
&&I_1(1,1,1)=\frac{{\rm i}}{192 \pi ^2},\hspace{1.5cm}I_2(1,1,1)=\frac{{\rm i}}{192 \pi ^2},\hspace{1.4cm}I_3(1,1,1,1)=-\frac{{\rm i}}{320 \pi ^2},
\nonumber\\&&I_4(1,1,1,1)=-\frac{{\rm i}}{480 \pi ^2},\hspace{0.8cm}I_5(1,1,1,1)=\frac{{\rm i}}{480 \pi ^2},\hspace{1.0cm}I_6(1,1,1,1)=\frac{{\rm i}}{320 \pi
^2},
\nonumber\\&&I_7(1,1,1,1)=-\frac{{\rm i}}{480 \pi ^2},\hspace{0.8cm}I_8(1,1,1)=-\frac{{\rm i}}{192 \pi ^2},\hspace{1.1cm}I_9(1,1,1)=-\frac{{\rm i}}{64 \pi ^2},
\nonumber\\&&I_{10}(1,1,1)=\frac{{\rm i}}{192 \pi ^2},\hspace{1.4cm}I_{11}(1,1,1)=-\frac{{\rm i}}{48 \pi ^2},
%%%%%%%%%%%%%%%%%%%%%%%%%%%%%%%%%%%%%%%%%
\\&&\hspace{0.01cm}\Delta^{LL}_{ij}(\tilde{D})=M^2_{SUSY}\delta^{LL}_{ij}(\tilde{D}),
\hspace{0.2cm}\Delta^{RR}_{ij}(\tilde{D})=M^2_{SUSY}\delta^{RR}_{ij}(\tilde{D}),
\hspace{0.2cm}\Delta^{LL}_{ij}(\tilde{U})=M^2_{SUSY}\delta^{LL}_{ij}(\tilde{U}),
\nonumber\\&&{T_d}^{ij}=M_{SUSY}\delta^{ij}_{T_d},
\hspace{1.8cm}{T_u}^{ij}=M_{SUSY}\delta^{ij}_{T_u}.\label{delta}
\end{eqnarray}
To simplify the study, we suppose that the used matrices are symmetric, for example ${T_d}^{ij}={T_d}^{ji}$.

Then, we obtain the much simplified one-loop results of $C_7$ and $C_8$
\begin{eqnarray}
&&C_7= \frac{{\rm i}\mathcal{A}g_3^2}{2304\pi^2 M_{SUSY}^2}\Big\{\frac{  \tan{\beta}}{15}\delta^{RR}_{23}(\tilde{D}) \Big(( 3 g_X^2 + g_X g_{YX} - 2
g_{YX}^2)\texttt{sign}[M_{\lambda_{\tilde{X}}} \mu_H'] - 2 g_1^2 \texttt{sign}[M_1 \mu_H'] \Big)
\nonumber\\&&+ \frac{\tan{\beta}}{15}\frac{m_s }{m_b }\delta^{LL}_{23}(\tilde{D})
\Big(3 g_2^2\texttt{sign}[M_2 \mu_H'] - (g_X g_{YX} + g_{YX}^2)\texttt{sign}[M_{\lambda_{\tilde{X}}} \mu_H'] - g_1^2 \texttt{sign}[M_1
\mu_H']\Big)
\nonumber\\&& + \frac{  v \cos{\beta}\delta^{23}_{T_d} }{180 m_b } \Big( 10 g_1^2 \texttt{sign}[M_1] + (15 g_X g_{YX}+10 g_{YX}^2 )
\texttt{sign}[M_{\lambda_{\tilde{X}}}]
- 8 g_1 g_{YX} \texttt{sign}[M_{\lambda_{\tilde{X}}} M_1 M_{BB'} ]
\nonumber\\&& - 45 g_3^2 \texttt{sign}[M_{\Lambda_G}] \Big)
 + \frac{\sqrt{2}M_{SUSY}}{v}\frac{ m_s }{m_b}g_2
 \Big(2 \tan{\beta} \texttt{sign}[\mu_H'] - \texttt{sign}[M_2]\Big)\sum_{l,n=1}^3 V_{2n}^* V_{3l} \delta^{LL}_{ln}(\tilde{U})
\nonumber\\&& - \frac{22m_s \texttt{sign}[\mu_H']}{ 5 m_b  v \cos{\beta}} \sum_{i,j=1}^3  V^*_{2i} V_{3j} m_{u_j} \delta^{ji}_{T_u}
\Big\},\label{C7}
%%%%%%%%%%%%%%%%%%%%%%%%%%%%%%%%%%%%%%%%%%%%%%%%%%%%%%
\\&&C_8=  \frac{{\rm i}\mathcal{A}g_3^3}{240\pi^2 M_{SUSY}^2}\Big\{\frac{\tan{\beta} }{12} \delta^{RR}_{23}(\tilde{D})\Big(( 3 g_X^2 + g_X g_{YX} - 2
g_{YX}^2)\texttt{sign}[M_{\lambda_{\tilde{X}}} \mu_H'] - 2 g_1^2 \texttt{sign}[M_1 \mu_H'] \Big)
\nonumber\\&&+ \frac{\tan{\beta} }{12} \frac{ m_s }{m_b }\delta^{LL}_{23}(\tilde{D}) \Big(3 g_2^2 \texttt{sign}[M_2 \mu_H'] - (g_X g_{YX} +
g_{YX}^2)\texttt{sign}[M_{\lambda_{\tilde{X}}} \mu_H'] - g_1^2 \texttt{sign}[M_1 \mu_H']\Big)
\nonumber\\&& + \frac{ v \cos{\beta} \delta^{23}_{T_d} }{144 m_b } \Big(10 g_1^2 \texttt{sign}[M_1] + (15 g_X g_{YX} + 10 g_{YX}^2 )
\texttt{sign}[M_{\lambda_{\tilde{X}}}]- 8  g_1 g_{YX} \texttt{sign}[M_{\lambda_{\tilde{X}}} M_1 M_{BB'}]
\nonumber\\&& - 60 g_3^2 \texttt{sign}[M_{\Lambda_G}]\Big) + \frac{\sqrt{2}M_{SUSY}}{4 v} \frac{m_s}{m_b} g_2  \Big(\texttt{sign}[M_2]- 4
\tan{\beta} \texttt{sign}[\mu_H']\Big)\sum_{l,n=1}^3 V^*_{2n} V_{3l} \delta^{LL}_{ln}(\tilde{U})
\nonumber\\&& - \frac{2 m_s\texttt{sign}[\mu_H']}{ m_b v \cos{\beta}} \sum_{i,j=1}^3  V^*_{2i} V_{3j} m_{u_j} \delta^{ji}_{T_u}
\Big\}.\label{C8}
\end{eqnarray}
It can be found that
$\texttt{sign}[M_1],~\texttt{sign}[M_2],~\texttt{sign}[M_{\lambda_{\tilde{X}}}],~\texttt{sign}[\mu_H^{\prime}],~\texttt{sign}[M_G]$ and
$\texttt{sign}[M_{BB^{\prime}}]$ have a certain impact on the corrections of $C_7$ and $C_8$.
According to  $1>g_X>g_{YX}>0$, we assume
$\texttt{sign}[M_1]=\texttt{sign}[M_{\lambda_{\tilde{X}}}]=\texttt{sign}[\mu_H^{\prime}]=\texttt{sign}[M_2]=\texttt{sign}[M_G]=\texttt{sign}[M_{BB^{\prime}}]=1$,
and get the larger values of $C_7$ and $C_8$
\begin{eqnarray}
&&C_7= \frac{{\rm i}\mathcal{A}g_3^2}{2304\pi^2 M_{SUSY}^2}\Big\{\frac{ \tan{\beta} } {15}\delta^{RR}_{23}(\tilde{D})
( 3 g_X^2 + g_X g_{YX} - 2 g_{YX}^2 - 2 g_1^2 )
+ \frac{ \tan{\beta} }{15 }\frac{m_s}{m_b}\delta^{LL}_{23}(\tilde{D})\nonumber\\&&\times
(3 g_2^2 - g_X g_{YX} - g_{YX}^2 - g_1^2 )
+ \frac{v \cos{\beta} \delta^{23}_{T_d} }{180  m_b } \big( 10 g_1^2 + 15 g_X g_{YX}+10 g_{YX}^2 - 8 g_1 g_{YX} - 45 g_3^2\big)
\nonumber\\&& + \frac{\sqrt{2}m_s M_{SUSY}}{  m_b  v}g_2 (2 \tan{\beta} - 1)\sum_{l,n=1}^3 V_{2n}^* V_{3l} \delta^{LL}_{ln}(\tilde{U})
 - \frac{22m_s}{5 m_b  v \cos{\beta}} \sum_{i,j=1}^3  V_{2i}^* V_{3j} m_{u_j} \delta^{ji}_{T_u}\Big\},\label{CC7}
\end{eqnarray}
\begin{eqnarray}
&&C_8= \frac{{\rm i}\mathcal{A}g_3^3}{240\pi^2 M_{SUSY}^2}\Big\{\frac{ \tan{\beta}}{12  }\delta^{RR}_{23}(\tilde{D})  ( 3 g_X^2 + g_X g_{YX} - 2
g_{YX}^2 - 2 g_1^2)
+ \frac{  \tan{\beta}}{12 } \frac{m_s}{m_b} \delta^{LL}_{23}(\tilde{D})\nonumber\\&&\times(3 g_2^2 -g_X g_{YX} - g_{YX}^2- g_1^2)
+ \frac{  v \cos{\beta} \delta^{23}_{T_d}}{144 m_b } (10 g_1^2 + 15 g_X g_{YX} + 10 g_{YX}^2 - 8  g_1 g_{YX} - 60 g_3^2)
\nonumber\\&& + \frac{\sqrt{2} m_s M_{SUSY} }{4 m_b  v}g_2 (1 - 4 \tan{\beta})\sum_{l,n=1}^3 V_{2n}^* V_{3l} \delta^{LL}_{ln}(\tilde{U})-
\frac{2 m_s }{ m_b  v \cos{\beta}} \sum_{i,j=1}^3  V_{2i}^* V_{3j} m_{u_j} \delta^{ji}_{T_u}\Big\}.\label{CC8}
\end{eqnarray}

To determine the sensitive parameters affecting the branching ratio more clearly and intuitively, similar method is used to derive
$\tilde{C}_{7}$ and $\tilde{C}_{8}$ in Eq.(\ref{BRNP}).
Then in combination with the degenerate results in Eq.(\ref{CC7}), Eq.(\ref{CC8}) and
Eq.(\ref{delta}), we set
$M_{SUSY}=\mu_H',~g_{YX}=0.1,~g_X=0.5,~\Delta^{RR}_{23}(\tilde{D})=\Delta^{LL}_{23}(\tilde{D})=1\times10^{4}~{\rm{GeV}^2},
~\Delta^{LL}_{ij}(\tilde{U})=4.2\times10^{4}~{\rm{GeV}^2}(i\ne j ~{\rm and}~ i,j=1,2,3)$ and $\Delta^{LL}_{ii}(\tilde{U})=5\times10^6~{\rm
GeV^2}~(i=1,2,3)$ to obtain the following figures:
\begin{figure}[h]
\setlength{\unitlength}{4mm}
\centering
\includegraphics[width=3.4in]{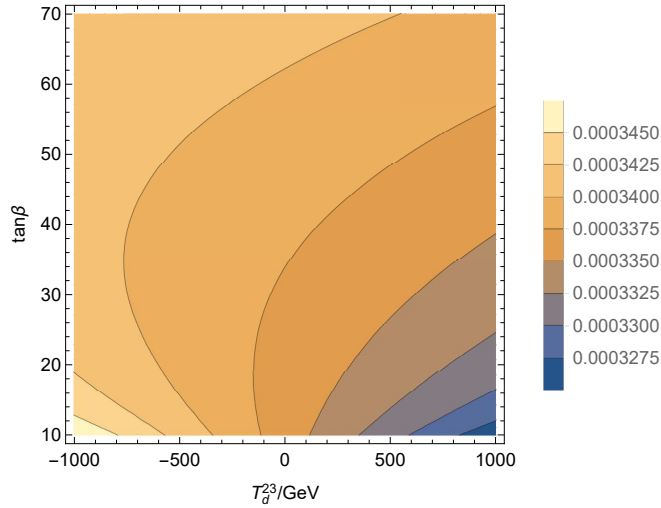}
\vspace{0.2cm}
\caption{With ${T_u}^{23}=-650~{\rm GeV}$ and $\mu=1000~{\rm GeV}$, the effects of $\tan{\beta}$ and ${T_d}^{23}$ on $Br(\bar{B}\rightarrow
X_s\gamma)$. The x-axis denotes the range of ${T_d}^{23}$ is from $-1000~{\rm GeV} $ to $ 1000~{\rm GeV} $, and the y-axis represents
$10<\tan\beta<70$. The different colors of the rightmost icon correspond to the values of $Br(\bar{B}\rightarrow X_s\gamma)$.}{\label {N3}}
\end{figure}
\begin{figure}[h]
\setlength{\unitlength}{5mm}
\centering
\includegraphics[width=3.4in]{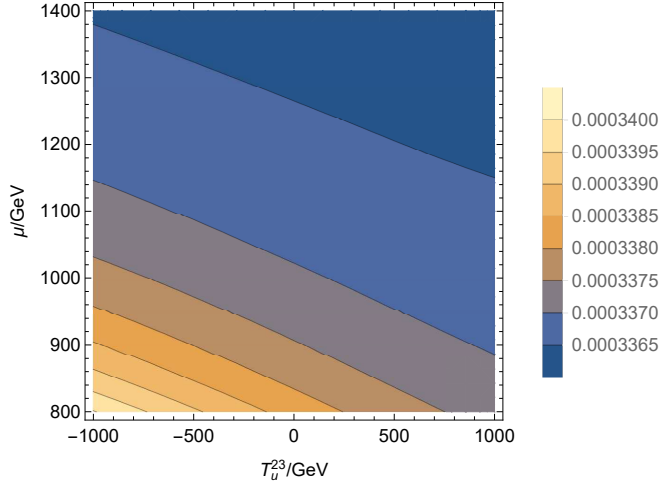}
\vspace{0.2cm}
\caption{With ${T_d}^{23}=600~{\rm GeV}$ and $\tan\beta=50$, the effects of $\mu$ and ${T_u}^{23}$ on $Br(\bar{B}\rightarrow X_s\gamma)$. The
x-axis denotes the range of ${T_u}^{23}$ is from $-1000~{\rm GeV}$ to $1000~{\rm GeV}$, and the y-axis represents $800~{\rm GeV}<\mu<1400~{\rm
GeV}$. The different colors of the rightmost icon correspond to the values of $Br(\bar{B}\rightarrow X_s\gamma)$.}{\label {N4}}
\end{figure}

Combined with Fig.~\ref{N3} and Fig.~\ref{N4}, it can be seen that $\tan{\beta}$, $\mu$, ${T_d}^{23}$ and ${T_u}^{23}$ have a significant
impact on the results. In Fig.~\ref{N3}, we can see that $Br(\bar{B}\rightarrow X_s\gamma)$ increases as the $\tan{\beta}$ increases. The
reason can be found in Eq.(\ref{CC7}) and Eq.(\ref{CC8}). $\tan{\beta}$ is almost always in the numerator, so it is proportional to the result.
When $\tan\beta=40$ and ${T_d}^{23}=-600~{\rm GeV}$, $Br(\bar{B}\rightarrow X_s\gamma)$ reaches $3.4\times10^{-4}$. Compared to the
influence of ${T_d}^{23}$, $\tan\beta$ obviously has stronger influence. In our hypothesis, $M_{SUSY}$ is equal to $\mu_H'$, and $\mu$ is
proportional to $\mu_H'$. In Eq.(\ref{CC7}) and Eq.(\ref{CC8}), $M_{SUSY}$ is inversely proportional to the result. Thus, $\mu$ should be
inversely proportional to the result, which can be demonstrated in Fig.~\ref{N4}. When $\mu=800~{\rm GeV}$ and ${T_u}^{23}=-730~{\rm GeV}$, the
branching ratio reaches $3.4\times10^{-4}$. And the influence of $\mu$ is greater than that of ${T_u}^{23}$. At the same time, as the values of
FCNC sources $\Delta^{LL}_{ij}(\tilde{D}),~\Delta^{RR}_{ij}(\tilde{D})$ and $\Delta^{LL}_{ij}(\tilde{U})$ go up, the result also grows. But
they have gentle influence on the result, which is not shown here. In summary, we can get the sensitive parameters including $\tan{\beta}$,
$\mu$, ${T_d}^{23}$ and ${T_u}^{23}$.

\section{numerical results}
To study $Br(\bar{B}\rightarrow X_s\gamma)$, we consider the mass constraint for the $Z^\prime$ boson ($M_{Z^{\prime}}> 5.1$ TeV) \cite{Zp5d1}
from the latest Large Hadron Collider (LHC) data \cite{LHC1,LHC2,LHC3,LHC4,LHC5,LHC6,LHC7}. The constraints $M_{Z^\prime}/g_X\geq 6 ~{\rm TeV}$
and $\tan \beta_\eta< 1.5$ are also taken into account. According to the research in Ref. \cite{gluino}, we take the mass of gluino more than 2
TeV. The parameters are used to make the scalar lepton masses larger than 700 GeV, and chargino masses larger than 1100 GeV, and the scalar quark
masses greater than 2000 GeV. Ref. \cite{tan1} and Ref. \cite{tan2} discuss parameter space for $b\to s\gamma$ under various models, and in their data
analysis the value of $\tan\beta$ is relatively high. Considering that $U(1)_X$SSM has more parameters and
larger adjustment space of parameters, we tend to choose a large value of $\tan\beta$.

In this section, we discuss the numerical results of branching ratio with some assumptions. Other free parameters introduced in the $U(1)_X$SSM
are set to
\begin{eqnarray}
&&g_{X}=0.4,~g_{YX}=0.1,~\lambda_H=0.1,~v_S=3.5~{\rm TeV},~T_{u11}=T_{u22}=T_{u33}=3~{\rm TeV},
\nonumber\\&&M_1=1~{\rm TeV},~M_2=1.1~{\rm TeV},~M_{\lambda_{\tilde{X}}}=1.4~{\rm TeV},~M_G=2.1~{\rm TeV},~M_{BB'}=0.8~{\rm TeV}.
\end{eqnarray}

After roughly determining the sensitive parameters, in order to study the influence of parameters on $Br(\bar{B}\rightarrow X_s\gamma)$ better,
we also need to study before degenerate. To show the numerical results clearly, we draw the relation diagrams and scatter diagrams of
$Br(\bar{B}\rightarrow X_s\gamma)$ with the parameters of the $U(1)_X$SSM.
\begin{figure}[h]
\setlength{\unitlength}{5mm}
\centering
\includegraphics[width=4in]{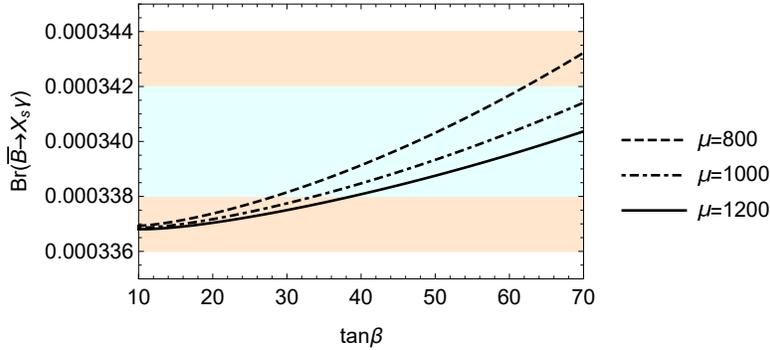}
\vspace{0.2cm}
\caption{$Br(\bar{B}\rightarrow X_s\gamma)$ versus $\tan{\beta}$ with $\Delta^{ij}=3\times10^5~{\rm GeV^2}$.}
\label{1W1}
\end{figure}
\begin{figure}[h]
\setlength{\unitlength}{5mm}
\centering
\includegraphics[width=4.5in]{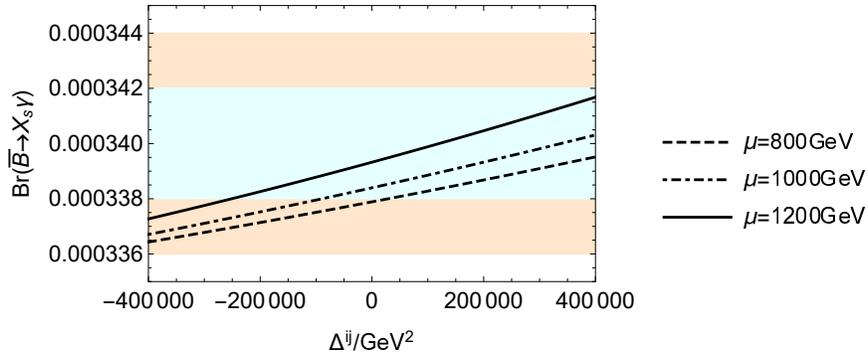}
\vspace{0.2cm}
\caption{$Br(\bar{B}\rightarrow X_s\gamma)$ versus $\Delta^{ij}$ with $\tan\beta=60$.}
\label{1W2}
\end{figure}
Fig.~\ref{1W1} shows the relationship between $\tan{\beta}$ and $Br(\bar{B}\rightarrow X_s\gamma)$, where $Br(\bar{B}\rightarrow X_s\gamma)$
increases with the enlarging $\tan{\beta}$. As the lines in the figure go from bottom to top, the value of $\mu$ decreases gradually.
Meanwhile, under the same $\tan{\beta}$, the smaller the $\mu$ is, the larger the value of $Br(\bar{B}\rightarrow X_s\gamma)$ is. This is the
same as the conclusion of degenerate results. When the value of $\mu$ decreases, the influence of $\tan{\beta}$ gradually becomes more
significant. Since the influences of some FCNC sources are small, we set $\Delta^{LL}_{ij}(\tilde{D}),~\Delta^{RR}_{ij}(\tilde{D})$ and
$\Delta^{LL}_{ij}(\tilde{U})$ to be equal in the drawing of the diagram, and their values are represented by $\Delta^{ij}(i\ne j)$, so as to
get Fig.~\ref{1W2}. The result is proportional to $\Delta^{ij}$. When $\Delta^{ij}=1.3\times10^5~{\rm GeV}^2$ and $\mu=800~{\rm GeV}$,
$Br(\bar{B}\rightarrow X_s\gamma)$ reaches $3.4\times10^{-4}$. Compared with the degenerate result, the influence degree of some parameters on
the result has changed. For example, the effects of ${T_d}^{23}$ and ${T_u}^{23}$ are reduced in precise results. But $\Delta^{ij}$, which have
less effect on the degenerate result, have more effect here.

\begin{figure}[ht]
\setlength{\unitlength}{5mm}
\centering
\includegraphics[width=3in]{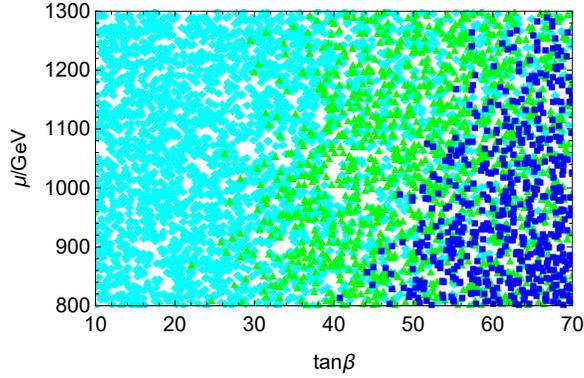}
\vspace{0.2cm}
\caption{The effects of $\tan\beta$ and $\mu$ on $Br(\bar{B}\rightarrow X_s\gamma)$.}
\label{2W1}
\end{figure}

\begin{figure}[h]
\centering
\subfigure[]{
\setlength{\unitlength}{5mm}
\includegraphics[width=3in]{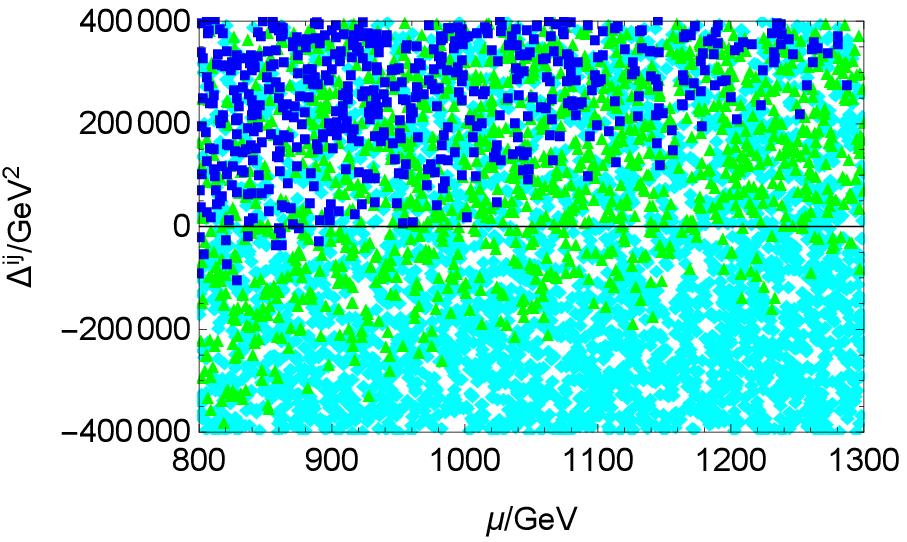}
\label{2W2}
}
\subfigure[]{
\setlength{\unitlength}{5mm}
\includegraphics[width=3in]{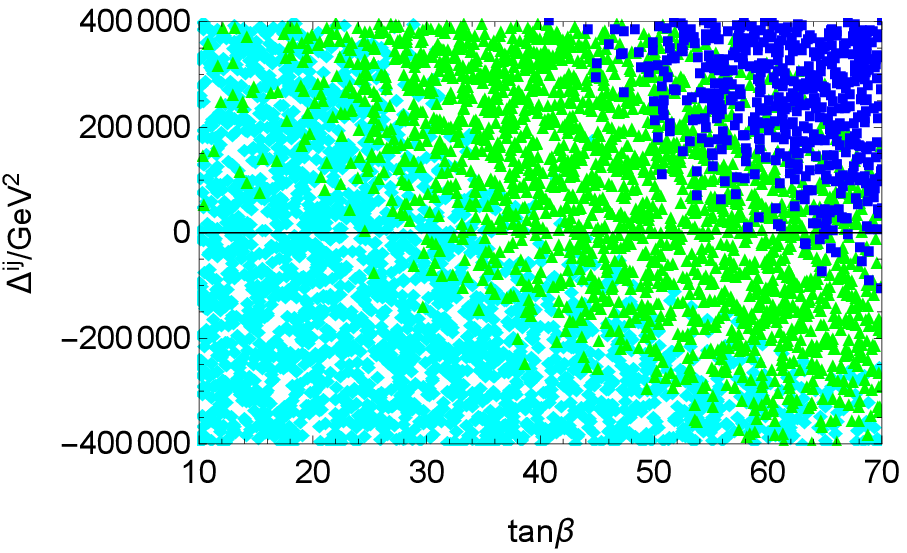}
\label{2W3}
}
\caption{The effects of $\mu$ and $\Delta^{ij}$ on $Br(\bar{B}\rightarrow X_s\gamma)$(a);~the effects of $\tan\beta$ and $\Delta^{ij}$ on
$Br(\bar{B}\rightarrow X_s\gamma)(b).$}
\label{2W}
\end{figure}

\begin{figure}[ht]
\setlength{\unitlength}{5mm}
\centering
\includegraphics[width=3in]{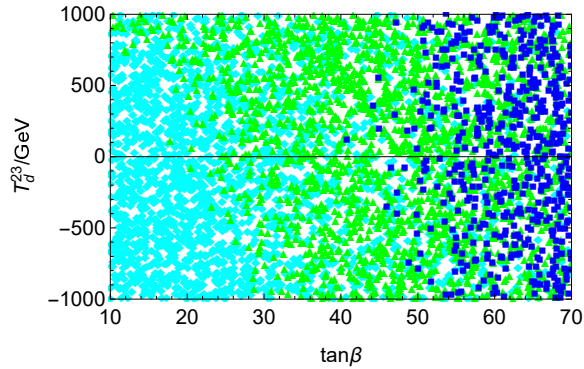}
\vspace{0.2cm}
\caption{The effect of $\tan\beta$ and $T_d^{23}$ on $Br(\bar{B}\rightarrow X_s\gamma)$.}
\label{2W4}
\end{figure}

For more multidimensional analysis of sensitive parameters, we draw the scatter points in Figs.~\ref{2W1}~-~\ref{2W4}. Under the
premise of current limit on flavor transition process $b\rightarrow s\gamma$, we select the parameter ranges as follows:
\begin{eqnarray}
&&\tan{\beta}=10\sim70,\hspace{2.4cm}{T_d}^{23}=-1000\sim1000~{\rm GeV},
\nonumber\\&&\mu=800\sim1300~{\rm GeV},\hspace{1.4cm}\Delta^{ij}=(-4.0\sim4.0)\times10^5~{\rm {GeV}^2}.
\end{eqnarray}
In Figs.~\ref{2W1}~-~\ref{2W4}, \textcolor{bright-blue} {$\blacklozenge$} mean the value of $Br(\bar{B}\rightarrow X_s\gamma)$ less than
$3.37\times10^{-4}$, \textcolor{green}{$\blacktriangle$} mean $Br(\bar{B}\rightarrow X_s\gamma)$ in the range of $3.37\times10^{-4}$ to
$3.39\times10^{-4}$, \textcolor{blue} {$\blacksquare$} show $3.39\times10^{-4}$ $\leq Br(\bar{B}\rightarrow X_s\gamma)$.

Fig.~\ref{2W1} shows the relationship between $\tan\beta$ and $\mu$, and it can be seen that the largest area is occupied by \textcolor{bright-blue} {$\blacklozenge$}. \textcolor{blue} {$\blacksquare$} are mainly concentrated on the right side of the image and decrease with the increase of
$\mu$ or the decrease of $\tan\beta$. When $\tan\beta$ is greater than 45, the result is easier to get to $3.40\times10^{-4}$. Moreover, the
three markers \textcolor{bright-blue} {$\blacklozenge$}, \textcolor{green}{$\blacktriangle$} and \textcolor{blue} {$\blacksquare$} have clear
distribution boundaries, and $\tan\beta$ has a greater impact on the results.

Fig.~\ref{2W2} shows the relationship between $\mu$ and $\Delta^{ij}$, and Fig.~\ref{2W3} is the relationship between $\tan\beta$ and
$\Delta^{ij}$. It can be further seen from the figure that $\tan\beta$ has more obvious influence than $\mu$, because the distribution of
midpoints in Fig.~\ref{2W2} is not as clear as that in Fig.~\ref{2W3}. For example marker \textcolor{bright-blue} {$\blacklozenge$}, in
Fig.~\ref{2W2}, marker \textcolor{bright-blue} {$\blacklozenge$} are widely distributed. In contrast, the markers in Fig.~\ref{2W3} are mainly
concentrated in the lower left part, and there is no marker \textcolor{bright-blue} {$\blacklozenge$} in the upper right part. Combining the two graphs,
$\Delta^{ij}$ increase and so does the branching ratio. And when $\Delta^{ij}$ are greater than $-1\times10^{-5}~{\rm GeV^2}$, it is easier to
reach $3.40\times10^{-4}$.
The relationship between ${T_d}^{23}$ and $\tan\beta$ is shown in Fig.~\ref{2W4}.
It can be seen that the results increase with the enlarging
${T_d}^{23}$. However, when $\tan\beta$ increases to a certain extent,
the influence of ${T_d}^{23}$ turns weak gradually.

\section{discussion and conclusion}
The $U(1)_X$SSM has new superfields including right-handed neutrinos $\hat{\nu}_i$ and three Higgs superfields
$\hat{\eta},~\hat{\bar{\eta}},~\hat{S}$, and its local gauge group is $SU(3)_C\otimes SU(2)_L \otimes U(1)_Y\otimes U(1)_X$. As an interesting process of FCNC, the flavor transition process $b\rightarrow s\gamma$ is investigated by the MIA within the framework of the $U(1)_X$SSM. With effective Hamiltonian method, we present the Wilson coefficients extracted from amplitudes corresponding to the concerned one-loop diagrams. Based on the analytical results, constraints on the parameters are given with the experimental data of $Br(\bar{B}\rightarrow X_s\gamma)$. From the data analysis, the flavor transition process $b\rightarrow s\gamma$ has relatively stricter restrictions on the quark flavor violation sources than most other parameters.

We take into account the constraint from the branching ratio. By MIA, we obtain simpler degenerate results. These results contribute to us to more intuitively determine sensitive parameters. It is convenient for subsequent numerical analysis. Through the multi-angle analysis of the values, it is found that the branching ratio is more dependent on the variables $\tan{\beta}$ and $\mu$ in the $U(1)_X$SSM. $\Delta^{ij}$ are composed of quark flavor violation sources $\Delta^{LL}_{ij}(\tilde{D}), ~\Delta^{RR}_{ij}(\tilde{D})$ and $\Delta^{LL}_{ij}(\tilde{U})$, and the influence is also obvious. The effects of ${T_d}^{23}$ and ${T_u}^{23}$ are relatively insignificant. Parameters that are not graphed, such as $g_X$ and $g_{YX}$, have gentle effects on the results. $g_X,~g_{YX},~g_1,~ g_2,$ and $g_3$ are the coupling constants. $g_X$ and $g_{YX}$ only appear in some part of analytical result. So the impact of $g_X$ and $g_{YX}$ is relatively small. Therefore, we can get the main sensitive parameters that are $\tan\beta$, $\mu$ and the FCNC sources $\Delta^{ij}$. This work is conductive to further research on the $U(1)_X$SSM and FCNC.

{\bf Acknowledgments}

This work is supported by National Natural Science Foundation of China (NNSFC)
(No. 12075074), Natural Science Foundation of Hebei Province
(A2020201002, A202201022, A2022201017), Natural Science Foundation of Hebei Education
Department (QN2022173), Post-graduate's Innovation Fund Project of Hebei University
(HBU2023SS043), the youth top-notch talent support program of the Hebei Province.

{\bf Data Availability Statement:} The data are available from the corresponding author
on reasonable request.

\end{document}